\documentclass[twocolumn,showpacs,preprintnumbers,amsmath,amssymb]{revtex4}
\usepackage{graphicx}% Include figure files
\usepackage{dcolumn}% Align table columns on decimal point
\usepackage{bm}% bold math
\usepackage{amssymb}
\usepackage{amssymb}

\def\>{\rangle}

\newcommand{\ucite}[1]{$^{\mbox{\scriptsize \cite{#1}}}$}
\newtheorem{theorem}{Theorem}

\newtheorem{appendixlemma}{Lemma}[section]

\begin{document}

\title{Asymptotically noise decoupling for Markovian open quantum systems}% Force line breaks with \\

\author{Jing Zhang$^1$}\email{zhangjing97@mails.tsinghua.edu.cn}
\author{Re-Bing Wu$^2$}
\author{Chun-Wen Li$^1$}
\author{Tzyh-Jong Tarn$^3$}
\author{Jian-Wu Wu$^1$}
 \affiliation{%
$^1$ Department of Automation, Tsinghua University, Beijing
100084, P. R. China\\
$^2$ Department of Chemistry, Princeton University, Princeton, NJ
08544, USA\\
$^3$ Department of Electrical and Systems Engineering, Washington
University, St. Louis, MO 63130, USA
}%Lines break automatically or can be forced with \\

\date{\today}% It is always \today, today,
             %  but any date may be explicitly specified

\begin{abstract}
The noise decoupling problem is investigated for general N-level
Markovian open quantum systems. Firstly, the concept of Cartan
decomposition of the Lie algebra $su(N)$ is introduced as a tool
of designing control Hamiltonians. Next, under certain
assumptions, it is shown that a part of variables of the coherence
vector of the system density matrix can be asymptotically
decoupled from the environmental noises. The resulting noise
decoupling scheme is applied to one-qubit, qutrit and two-qubit
quantum systems, by which the coherence evolution of the one-qubit
and qutrit systems can always be asymptotically preserved, while,
for two-qubit systems, our findings indicate that evolution of
some variables can be preserved only for some initial states.
\end{abstract}

\pacs{03.67.Lx,03.67.Mn,03.67.Pp}% PACS, the Physics and Astronomy
                             % Classification Scheme.
%\keywords{Suggested keywords}%Use showkeys class option if keyword
                              %display desired
\maketitle

\section{Introduction}
\label{s1}

In recent years, quantum information science
has\ucite{Neilsen,Shor,Grover,Knill,Nakamura,Yamamoto,Wurebing}
been a growing field which interests many researchers for
potential high speed quantum computation and high security quantum
communication. Decoherence is commonly recognized as the main
bottleneck. Various schemes have been proposed to reduce such
unexpected effects. In principle, there are two classes of
schemes---open loop and closed loop strategies depending on the
use of measurement and feedback. Open loop strategies include
quantum error-avoiding codes\ucite{Zanardi,Chuang,Duan1},
Bang-Bang control\ucite{Viola1,Viola2,Viola3,Qiao}, open loop
optimal control\ucite{Khaneja1,Alessandro,Zhangjing1} and open
loop coherent control\ucite{Altafini1,Lidar}, while closed loop
strategies include quantum error-correction
codes\ucite{Shor_Steane_Knill,Cirac,Zurek,Duan2} and quantum
feedback control\ucite{Tombesi_Goetsch_Vitali,Fortunato}.

Though various strategies have been proposed, none of these
strategies is satisfying to suppress decoherence for $N$-level
Markovian open quantum systems: quantum error-correction codes and
error-avoiding codes use several physical qubits to encode one
logical qubit, which is too luxurious under existing conditions;
BangBang control strategy is inapplicable in the fully Markovian
regime as pointed out by Lidar\ucite{Lidar}; open-loop optimal
control strategy can only partially decouple quantum systems from
the environmental noises; quantum feedback control strategy
requires complex feedback control apparatus and it is valid only
for special physical systems. Thus, for the decoherence
suppression problem of $N$-level Markovian open quantum systems,
more system analysis and control methods should be introduced.

The closest work to ours can be found in Ref. \cite{Lidar}, where
open loop coherent control is applied to decoherence suppression
for single-qubit Markovian systems. It is shown that the $x$-axis
and $y$-axis variables of the Bloch vector can be exactly
decoupled from the environmental noises, which means that the
coherence of the quantum state can be well preserved. However,
this scheme requires solving a time-variant linear ordinary
differential equation (ODE) to obtain the open loop control laws,
by which analytic control laws can be obtained only for phase
damping decoherence under a strong assumption that the $x$-axis
and $y$-axis variables of the Bloch vector keep constant. For more
general cases, only numerical control laws can be obtained.
Furthermore, divergence of the control fields may occur in this
strategy as pointed out by the authors\ucite{Lidar}. In this
paper, we propose a more general noise decoupling strategy for
$N$-level Markovian open quantum systems based on the Cartan
decomposition of the Lie algebra $su(N)$. Open loop controls are
designed to asymptotically decouple the state variables from the
environmental noises. The strategy loses some precision but is
easier to be fulfilled.

The paper is organized as follows. In section \ref{s2}, Markovian
open quantum systems are formulated in the coherence vector
representation and the concept of the Cartan decomposition of
$su(N)$, together with three important assumptions, is introduced.
Our main results are presented in section \ref{s3}, where the
scheme are also applied to one-qubit, qutrit and two-qubit
systems. Further discussion and conclusion are drawn in section
\ref{s4}.

\section{Preliminaries}\label{s2}

Consider an N-level Markovian open quantum control system in the
following master equation form:
\begin{eqnarray}\label{Control master equation}
&&\dot\rho=-i[H_0+\sum\limits_{i=1}^n
u_iH_i,\rho]+\sum\limits_{j=1}^m\Gamma_j \mathbb{D}[L_j]\rho, \nonumber\\
&&\rho(t_0)=\rho_0,
\end{eqnarray}
where the Planck constant $\hbar$ is assigned to be $1$; $\rho$
refers to the system density matrix; $H_0$ and
$H_i,\,i=1,\cdots,n$ represent, respectively, the free Hamiltonian
and the control Hamiltonians adjusted by the control parameters
$u_i,\,i=1,\cdots,n$. The Lindblad super operators
$$\mathbb{D}[L_j]\rho=L_j\rho L_j^\dagger -\frac{1}{2}L_j^\dagger
L_j\rho-\frac{1}{2}\rho L_j^\dagger L_j,$$ characterize the
damping channels and the positive constants $\Gamma_j$ denote the
damping rates of the corresponding channels.

The differential equation (\ref{Control master equation}) is
actually a complex matrix differential equation which is hard to
be analyzed. Therefore, we will convert it into a real vector
differential equation. For this purpose, an orthonormal basis
$\{\Omega_0=\frac{1}{\sqrt{N}} I,\Omega_j \}_{j=1,\cdots, N^2-1}$
with respect to the matrix inner product $\langle
X,Y\rangle=tr(X^\dagger Y)$ should be introduced first, where $I$
is the $N\times N$ identity matrix and $\Omega_j's$ are $N\times
N$ Hermitian traceless matrices. The system density matrix $\rho$
can then be expressed as:
\begin{equation}\label{Coherent vector of the system density matrix}
\rho=\frac{1}{N}I+\sum_{i=1}^{N^2-1} m_i
\Omega_i:=\frac{1}{N}I+m\cdot \vec{\Omega},
\end{equation}
where $m\in \mathbb{R}^{N^2-1}$ is the so-called coherence
vector\ucite{Zhangjing1,Altafini1,Lidar,Alicki,Altafini2,Altafini3}
of $\rho$. In this case, the quantum control system (\ref{Control
master equation}) can be reexpressed by a differential equation on
$\mathbb{R}^{N^2-1}$:
\begin{eqnarray}\label{Coherent vector control system}
&&\dot{m}(t)=O_0 m(t)+\sum\limits_{i=1}^n u_i O_i m(t)+D m(t)+g,
\nonumber\\
&&m(t_0)=m_0,
\end{eqnarray}
where $O_0, O_i\in so(N^2-1)$ are, respectively, the adjoint
representation matrices of $-i H_0,\,-i H_i$\ucite{Altafini2} and
$m_0$ is the coherence vector of $\rho_0$. The term ``$Dm+g$"
comes from the decohering process represented by the Lindblad
terms in (\ref{Control master equation}) in which the dissipative
matrix $D$ is semi-negative defined, i.e., $D\leq 0$.

The above approach is a generalization of the well-known Bloch
vector representation for two-level quantum systems. Physically,
the length of the coherence vector represents the amount of
coherence in the quantum state. Concretely, it is unit for the
pure state and shorter for the mixed state. Notice that the
control $\sum_i u_i H_i$ drives the quantum state along a sphere
on which coherence is conserved, while the decohering operators
pull the vector towards the equilibrium state, e.g. the ground
state in a spontaneous emission.

To simplify the equation (\ref{Coherent vector control system})
and facilitate our discussions, it is useful to discuss the choice
of the matrix basis $\{ \frac{1}{\sqrt{N}} I,\,\,\Omega_j
\}_{j=1,\cdots,N^2-1}$. In this regard, we introduce the so-called
{\bf Cartan decomposition}\ucite{Khaneja2,Zhangjun,Mandilara} of
the Lie algebra $su(N)$ as follows:
\begin{equation}\label{Noise decoupling decomposition of su(N)}
su(N)=p\oplus\epsilon,\,\,[\epsilon,\epsilon]\subset\epsilon,\,\,[p,p]\subset\epsilon,\,\,[p,\epsilon]\subset
p.
\end{equation}

Notice that $su(N)$ is a $N^2-1$ dimensional Lie algebra of all
traceless skew-Hermitian $N\times N$ matrices, hence the basis
matrices of $p$ and $\epsilon$ can be expressed as $\{-i
\Omega_i^p \}_{i=1,\cdots,m}$ and $\{ -i\Omega_l^{\epsilon}
\}_{l=m+1,\cdots,N^2-1}$ where $\Omega_i^p$ and
$\Omega_l^{\epsilon}$ are traceless Hermitian matrices. Further,
every traceless Hermitian matrix is a linear combination of $\{
\Omega_i^p,\,\,\Omega_l^{\epsilon} \}$. Correspondingly, the
density matrix $\rho$ can be represented as:
\begin{equation}\label{Noise decoupling decomposition representation of the system density matrix}
\rho=\frac{1}{N}I+\sum_{i=1}^m m_i^p
\Omega_i^p+\sum_{l=m+1}^{N^2-1} m_l^{\epsilon}
\Omega_l^{\epsilon}.
\end{equation}
As will be shown later, the matrices $\{ \Omega_i^p
\}_{i=1,\cdots,m}$ play central roles in our control strategy. In
fact, {\bf we will choose the control Hamiltonians $H_i$ from
these matrices}.

It should be pointed out that the Cartan decomposition always
exists. In fact, we can obtain a trivial decomposition if we let
$\epsilon=su(N)$. The decomposition is also not unique.

Before proceeding our discussions, we introduce three assumptions
for the equation (\ref{Coherent vector control system}):
\begin{enumerate}
  \item [(H1)] Complete decoherence condition:
  $$[O_0,D]=0,\quad O_0
  g=0;$$
  \item [(H2)] Convergence condition: $D<0$;
  \item [(H3)] $-iH_0\in\epsilon$.
\end{enumerate}

(H1) has an important physical interpretation\ucite{Zhangjing1}
that the stationary distribution $\rho_{\infty}$ of the
uncontrolled system satisfies $[H_0,\rho_{\infty}]=0$ (see the
appendix of Ref. \cite{Zhangjing1} for a rigorous proof). In other
words, in the energy representation, the off-diagonal entries of
the stationary system density matrix $\rho_{\infty}$ disappear as
a result of decoherence. (H2) is introduced to guarantee the
existence of the convergent solution of (\ref{Coherent vector
control system}). In fact, we can always make $D<0$ with the aid
of the feedback control modification\ucite{Zhangjing1}. (H3) is
easy to be satisfied under special choices of the Cartan
decomposition.

\section{Asymptotically noise decoupling strategy}\label{s3}
The term ``$Dm+g$" in equation (\ref{Coherent vector control
system}) is the environment-induced dissipative term which
destroys coherence of the quantum states. Our target is to select
the control Hamiltonians $H_i$ and design the corresponding
controls $u_i$ to force the trajectory $m(t)$ of the equation
(\ref{Coherent vector control system}) as close as possible to the
target trajectory $m^0(t)$ which is the solution of the following
unperturbed system:
\begin{equation}\label{Coherent vector free system}
\dot{m}(t)=O_0 m(t),\,\, m(t_0)=m_0.
\end{equation}

It has been demonstrated that the Markovian open control system
(\ref{Coherent vector control system}) is always
uncontrollable\ucite{Altafini2}. Therefore, one can never track
the target trajectory precisely. However, we will show that
certain variables of $m(t)$ may asymptotically tend to the
corresponding variables of $m^0(t)$ by properly designed control
laws. That is to say, these variables can be asymptotically
decoupled from the environmental noises.

In fact, according to (\ref{Noise decoupling decomposition
representation of the system density matrix}), the coherence
vector $m$ can be divided into two parts: $m=(m^1,m^2)^T$, where
\begin{eqnarray*}
m^1&=&(m_1^p,\cdots,m_m^p)^T,\\
m^2&=&(m_{m+1}^{\epsilon},\cdots,m_{N^2-1}^{\epsilon})^T,
\end{eqnarray*}
From lemma \ref{Lemma of the expression of O_0, O_i}, we have
$O_0=diag(O_0^{11},O_0^{22})$, where $O_0^{11},\,O_0^{22}$ are
respectively, $m$ and $N^2-m-1$ dimensional square anti-symmetric
matrices, then the target trajectory can be written as:
$$m^0(t)=e^{O_0(t-t_0)}m_0=(m^{10}(t),m^{20}(t))^T,$$
where
\begin{eqnarray*}
&&m^{10}(t)=e^{O_0^{11}(t-t_0)}m_0^1,\\
&&m^{20}(t)=e^{O_0^{22}(t-t_0)}m_0^2,
\end{eqnarray*}
and $m_0=(m_0^1,m_0^2)^T$. In this case, the vector $m^1(t)$ can
be driven to the corresponding target trajectory $m^{10}(t)$. In
fact, we have the following theorem:
\begin{theorem}\label{Noise decoupling control design}
Suppose the assumptions (H1), (H2) and (H3) are satisfied and the
control Hamiltonians $H_i$ in (\ref{Control master equation}) are
the basis matrices $\{ \Omega_i^p \}$ corresponding to $p$. The
following control law
\begin{equation}\label{Control law}
u=(u_1,\cdots,u_m)^T=e^{-O^{11}_0(t-t_0)}\xi
\end{equation}
steers the control trajectory $m(t)$ of the equation
(\ref{Coherent vector control system}) asymptotically to the
stationary solution:
\begin{equation}\label{Stationary solution}
m^{\infty}(t)=(e^{O_0^{11}(t-t_0)}m_0^1,e^{O_0^{22}(t-t_0)}\eta)^T,
\end{equation}
where the constant vectors $\xi\in\mathbb{R}^m$ and
$\eta\in\mathbb{R}^{N^2-m-1}$ are the solutions of the following
nonlinear algebraic equation:
\begin{eqnarray}\label{Stationary algebraic equation}
F_1(\xi,\eta)&=&\sum_{i=1}^m \xi_i O_i^{12} \eta+D_{11}m_0^1+D_{12}\eta+g_1=0,\nonumber\\
F_2(\xi,\eta)&=&-\sum_{i=1}^m \xi_i (O_i^{12})^T
m_0^1+D_{21}m_0^1\nonumber\\
&&+D_{22}\eta+g_2=0,
\end{eqnarray}
where
$$D=\left(%
\begin{array}{cc}
  D_{11} & D_{12} \\
  D_{21} & D_{22} \\
\end{array}%
\right),\quad g=\left(%
\begin{array}{c}
  g_1 \\
  g_2 \\
\end{array}%
\right).$$
\end{theorem}

Rigorous proof of the theorem is left to appendix \ref{Proof Of
the noise decoupling control design theorem}. Theorem \ref{Noise
decoupling control design} shows that the state variables of the
coherence vector of $\rho$ corresponding to $p$ can be
asymptotically decoupled from the environmental noises. Note that
$O^{11}_0$ is an $m$-dimensional anti-symmetric matrix which has
only zero or pure imaginary eigenvalues, {\bf the control laws
presented in (\ref{Control law}) are in fact sinusoidal signals}
which are usual for electro-magnetic fields used in the
laboratory.

Next, we will apply our control strategy to three typical systems
in quantum information science: one-qubit, qutrit and two-qubit
systems.

\subsection{One-qubit quantum systems}\label{s31}

Consider the one-qubit systems which are fundamental in the
quantum information science. Assume that the free Hamiltonian
$H_0=\omega\sigma_z$ where $\sigma_z$ is the $z$-axis Pauli
matrix, then the Cartan decomposition of the corresponding Lie
algebra $su(2)$ can be chosen as $su(2)=p\oplus\epsilon$, where
the basis of $p$ and $\epsilon$ are $\{
-i\frac{1}{\sqrt{2}}\sigma_x,-i\frac{1}{\sqrt{2}}\sigma_y \}$ and
$\{ -i\frac{1}{\sqrt{2}}\sigma_z \}$ respectively.

Now, for the coherence vector representation of $\rho$, i.e.,
$$\rho=\frac{1}{2}I+m_x\frac{1}{\sqrt{2}}\sigma_x+m_y\frac{1}{\sqrt{2}}\sigma_y+m_z\frac{1}{\sqrt{2}}\sigma_z,$$
we can conclude that the variables
$m_x=\frac{1}{\sqrt{2}}tr(\sigma_x\rho)$ and
$m_y=\frac{1}{\sqrt{2}}tr(\sigma_y\rho)$ can be asymptotically
decoupled from the environmental noises if the assumptions (H1),
(H2) and (H3) are satisfied.

As an example, we study the one-qubit amplitude damping
decoherence model\ucite{Neilsen} which can be used to describe
spontaneous emissions of the two-level atoms. In this case, the
control system can be expressed as the following master equation:
\begin{equation}\label{Master equation of the one-qubit amplitude damping decoherence model}
\dot{\rho}=-i[\omega\sigma_z+u_x\sigma_x+u_y\sigma_y,\rho]+\Gamma
D[\sigma_-]\rho,
\end{equation}
where $u_x,\,u_y$ are the controls and $\omega\in\mathbb{R}$
denotes the Rabi frequency; $\sigma_-=\sigma_x-i\sigma_y$ is the
lowering operator of the two-level system and $\Gamma>0$ denotes
the decoherence rate, e.g., the damping rate of the spontaneous
emission process. To simplify the calculations, we use the
well-known Bloch vector representation:
$$\rho=\frac{1}{2}(I+m_x\sigma_x+m_y\sigma_y+m_z\sigma_z),$$
where $m_x=tr(\sigma_x\rho)$, $m_y=tr(\sigma_y\rho)$ and
$m_z=tr(\sigma_z\rho)$. Note that this representation is different
from the coherence vector representation only by a trivial
constant multiplicative factor.

The master equation (\ref{Master equation of the one-qubit
amplitude damping decoherence model}) can be converted into the
following equation in the Bloch vector representation:
\begin{equation}\label{Coherence vector equation of the one-qubit amplitude damping decoherence model}
\dot{m}=\omega O_z m+u_x O_x m+u_y O_y m+Dm+g,
\end{equation}
where $m=(m_x,m_y,m_z)^T\in \mathbb{R}^3$ and
$$O_x=\left(%
\begin{array}{ccc}
  0 &  &  \\
   &  & -1 \\
   & 1 &  \\
\end{array}%
\right),\,O_y=\left(%
\begin{array}{ccc}
   &  & 1 \\
   & 0 &  \\
  -1 &  &  \\
\end{array}%
\right),$$
$$O_z=\left(%
\begin{array}{ccc}
   & -1 &  \\
  1 &  & \\
   &  & 0 \\
\end{array}%
\right)$$ are the basis elements of the Lie algebra $so(3)$ of the
3 dimensional orthogonal group;
$D=diag(-\frac{\Gamma}{2},-\frac{\Gamma}{2},-\Gamma)$ and
$g=(0,0,-\Gamma)^T$ come from the decoherence process. It can be
verified from (\ref{Master equation of the one-qubit amplitude
damping decoherence model}) and (\ref{Coherence vector equation of
the one-qubit amplitude damping decoherence model}) that the
assumptions (H1), (H2) and (H3) are all satisfied.

From the equation $[O_0,O_i^p]=\sum_j (O^{11}_0)_{ij} O_j^p$ and
$$[O_z,O_x]=O_y,\,[O_z,O_y]=-O_x,$$
it can be easily computed that $O^{11}_0=\left(%
\begin{array}{cc}
   & -\omega \\
  \omega &  \\
\end{array}%
\right)$. According to theorem \ref{Noise decoupling control
design}, the controls can be designed as:
\begin{eqnarray*}
\left(%
\begin{array}{c}
  u_x \\
  u_y \\
\end{array}%
\right)&=&e^{-O^{11}_0(t-t_0)}\xi\\
&=&\left(%
\begin{array}{c}
  \cos\omega(t-t_0)\xi_1-\sin\omega(t-t_0)\xi_2 \\
  \sin\omega(t-t_0)\xi_1+\cos\omega(t-t_0)\xi_2 \\
\end{array}%
\right)\\
&=&\left(%
\begin{array}{c}
  A\cos(\omega(t-t_0)+\phi) \\
  A\sin(\omega(t-t_0)+\phi) \\
\end{array}%
\right),
\end{eqnarray*}
under which $m_x,\,m_y$ tend to the corresponding variables of the
unperturbed system $\dot{m}=\omega O_z m$. The constant
$\xi=(\xi_1,\,\xi_2)^T$ can be solved by (\ref{Stationary
algebraic equation}) as follows:
\begin{eqnarray*}
\xi_1=\frac{\Gamma m_{0y}}{2C_0^2} \left(1\pm\sqrt{1-2C_0^2}\right),\\
\xi_2=\frac{-\Gamma m_{0x}}{2C_0^2}
\left(1\pm\sqrt{1-2C_0^2}\right),
\end{eqnarray*}
where $m_0=(m_{0x},m_{0y},m_{0z})^T$ is the initial state and
$C_0^2=m_{0x}^2+m_{0y}^2$ represents the initial coherence in the
quantum system. It can be observed that $C_0^2$ should be no
larger than $\frac{1}{2}$ to guarantee the existence of the
solution of (\ref{Stationary algebraic equation}). Amplitude and
phase of the control fields can then be calculated as:
\begin{eqnarray*}
A&=&\sqrt{\xi_1^2+\xi_2^2}=\frac{\Gamma}{2C_0}\left|1\pm\sqrt{1-2C_0^2}\right|,\\
\phi&=&arctg\frac{\xi_2}{\xi_1}=arctg\left(-\frac{m_{0x}}{m_{0y}}\right).
\end{eqnarray*}

\begin{figure}[h]
\centerline{
\includegraphics[width=1.7in,height=1.2in]{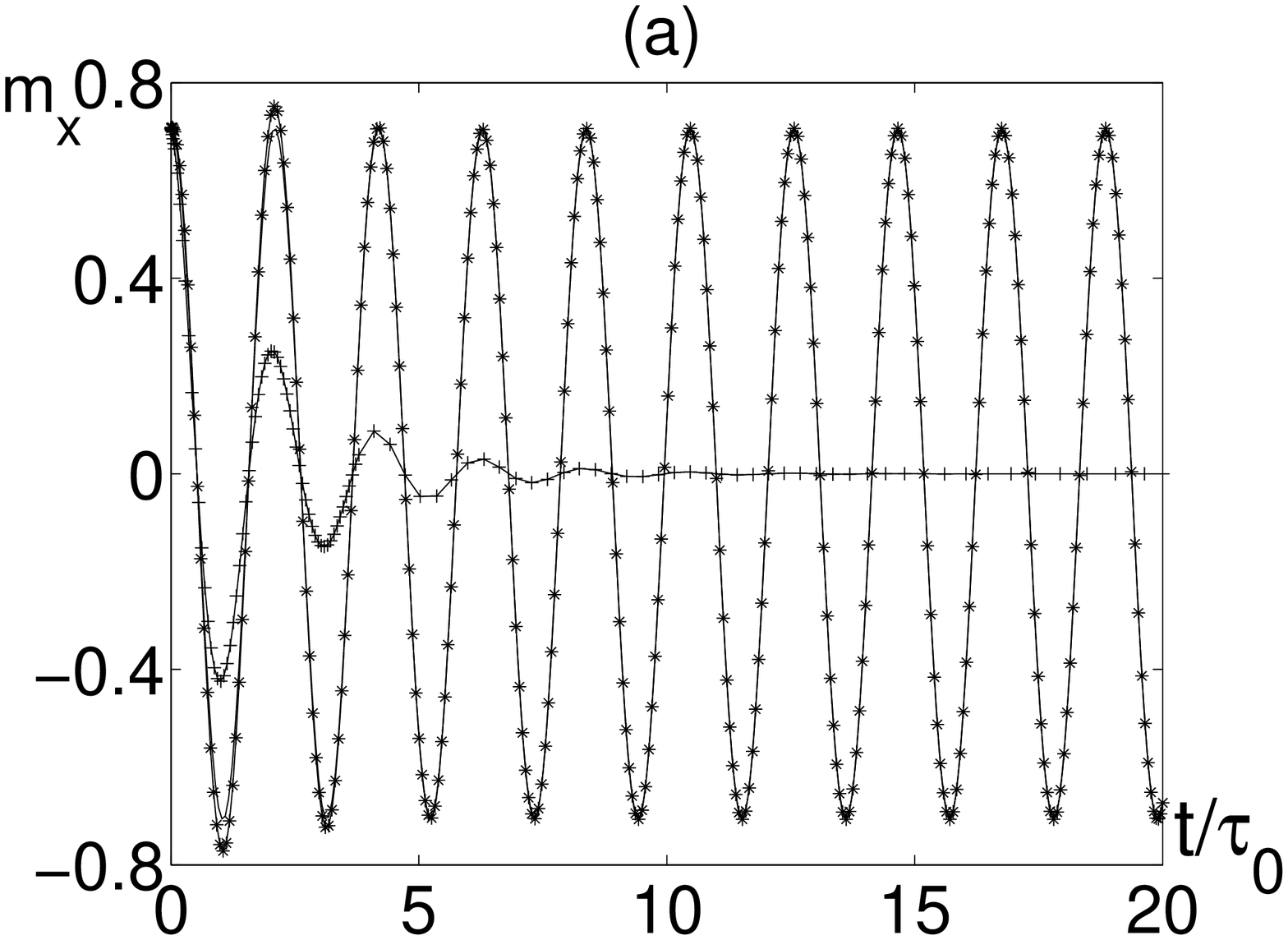}
\includegraphics[width=1.7in,height=1.2in]{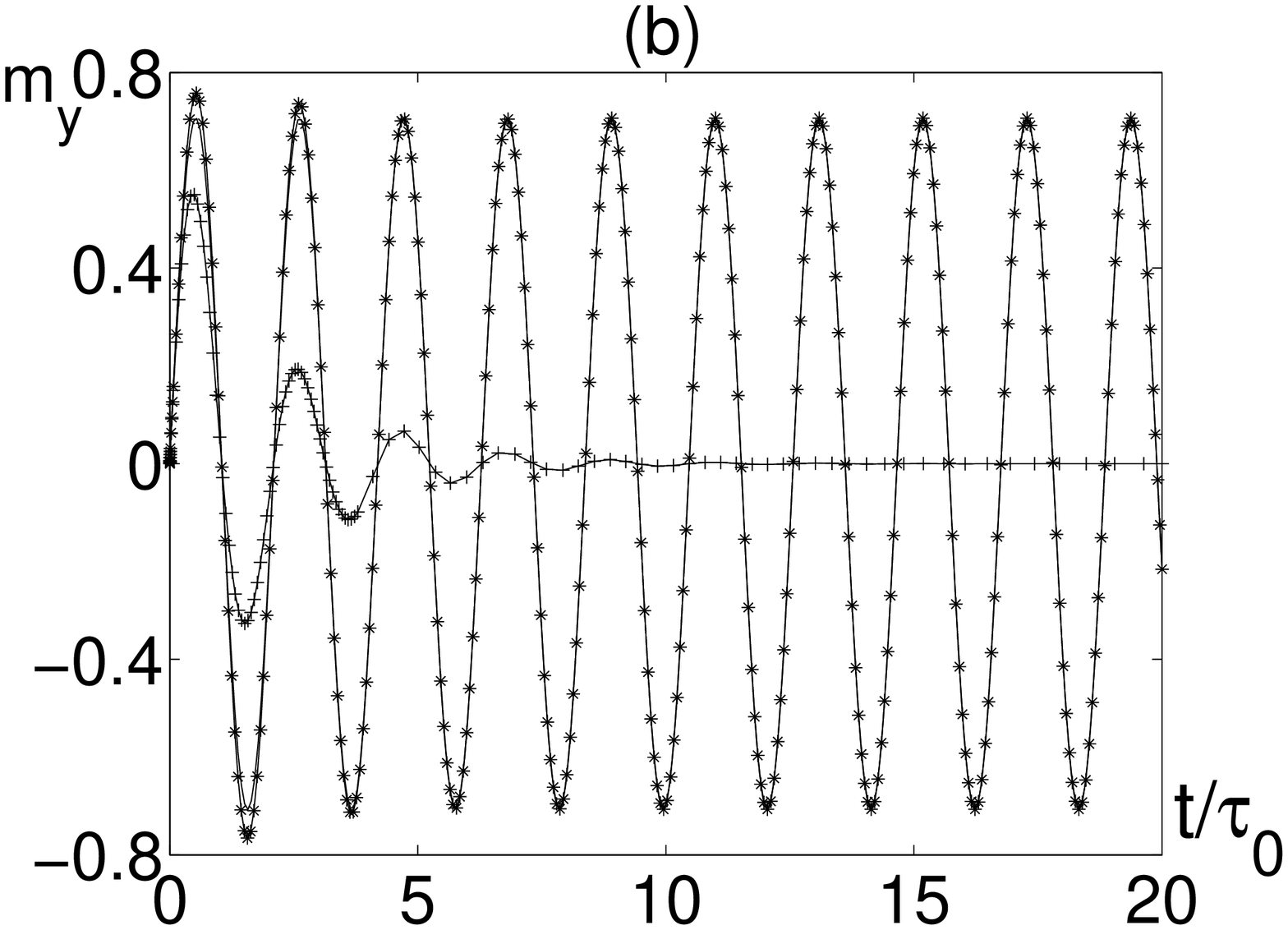}
} \caption{\scriptsize Temporal evolution of (a) $m_x$ and (b)
$m_y$: the asterisk line represents the controlled trajectory; the
plus-sign line is the uncontrolled trajectory; the solid line is
the target trajectory.}\label{Fig of the one-qubit systems}
\end{figure}

Let $\omega=3/{\tau_0}$, $t_0=0$ and the initial state
$$\rho_0=\frac{1}{2}I+\frac{\sqrt{2}}{4}\sigma_x+\frac{\sqrt{2}}{4}\sigma_z,$$
where $\tau_0$ is a time constant which is introduced to obtain
dimensionless evolution time. For concrete systems, $\tau_0$ can
be determined by the system time scale, e.g. relaxing time of
systems. With simple calculations, it can be shown that
$u_x=\frac{\sqrt{2}}{2}\Gamma\sin(3 t/{\tau_0})$,
$u_y=-\frac{\sqrt{2}}{2}\Gamma\cos(3 t/{\tau_0})$. Simulation
results of the variables $m_x$, $m_y$ and controls $u_x$, $u_y$
are shown in Figure {\ref{Fig of the one-qubit systems}} and
\ref{Fig of the one-qubit systems (controls)}.

\begin{figure}[h]
\centerline{
\includegraphics[width=2.5in,height=1.5in]{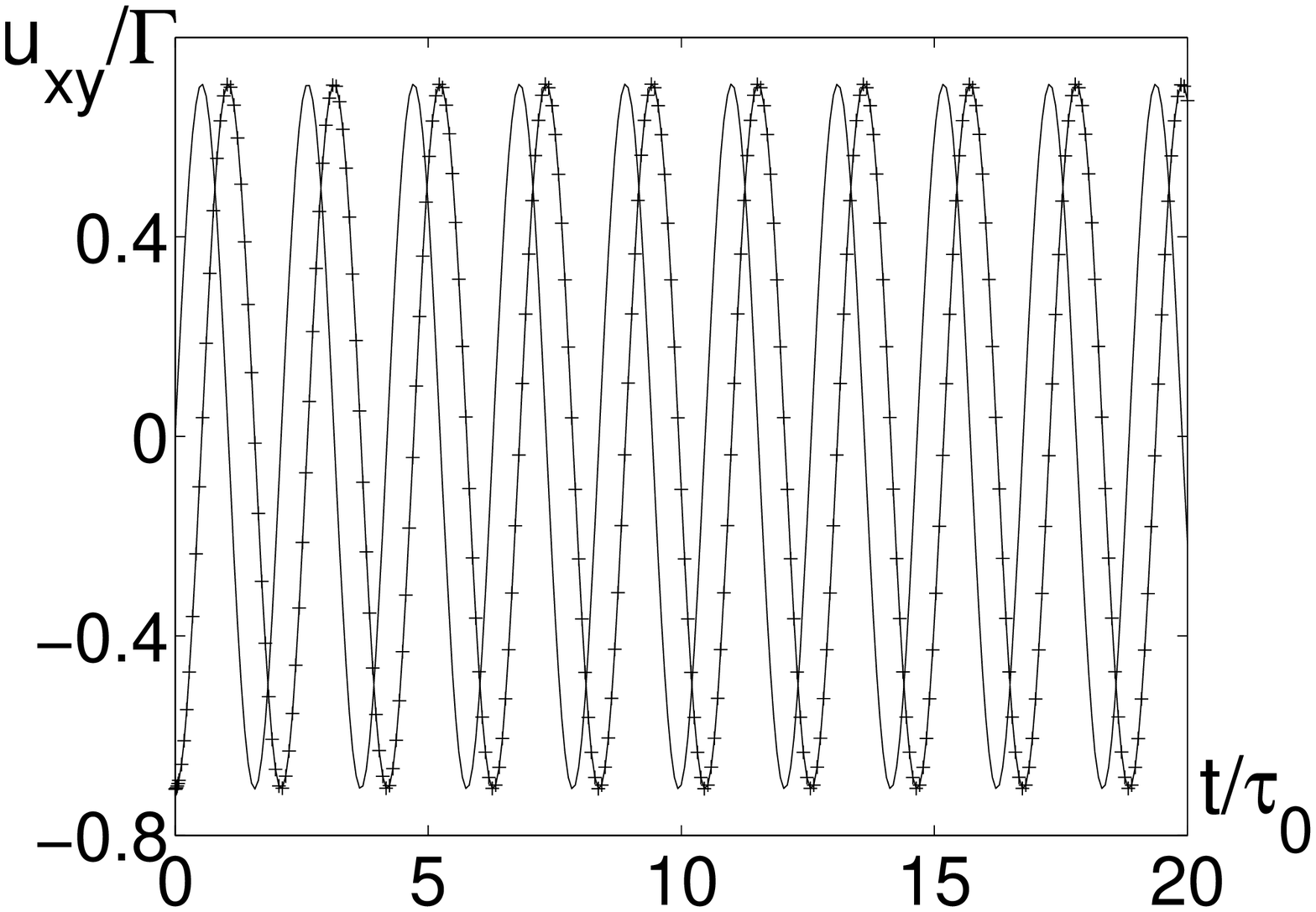}
} \caption{\scriptsize Temporal evolution of controls: the solid
line is $u_x$ and the plus-sign line is $u_y$. Here, to obtain
dimensionless quantities, $u_x$ and $u_y$ are divided by the
decoherence intensity $\Gamma$.}\label{Fig of the one-qubit
systems (controls)}
\end{figure}

Since the coherence $C^2=m_x^2+m_y^2$ in $\rho$ is determined by
$m_x$ and $m_y$, the coherence of the state will vanish completely
without control. The figure shows that the controlled trajectory
tends asymptotically to the target trajectory where the two
trajectories are so close that they almost coincide together,
which implies that the coherence of the state is asymptotically
preserved with our control strategy.

\begin{figure}[h]
\centerline{
\includegraphics[width=2.5in,height=1.5in]{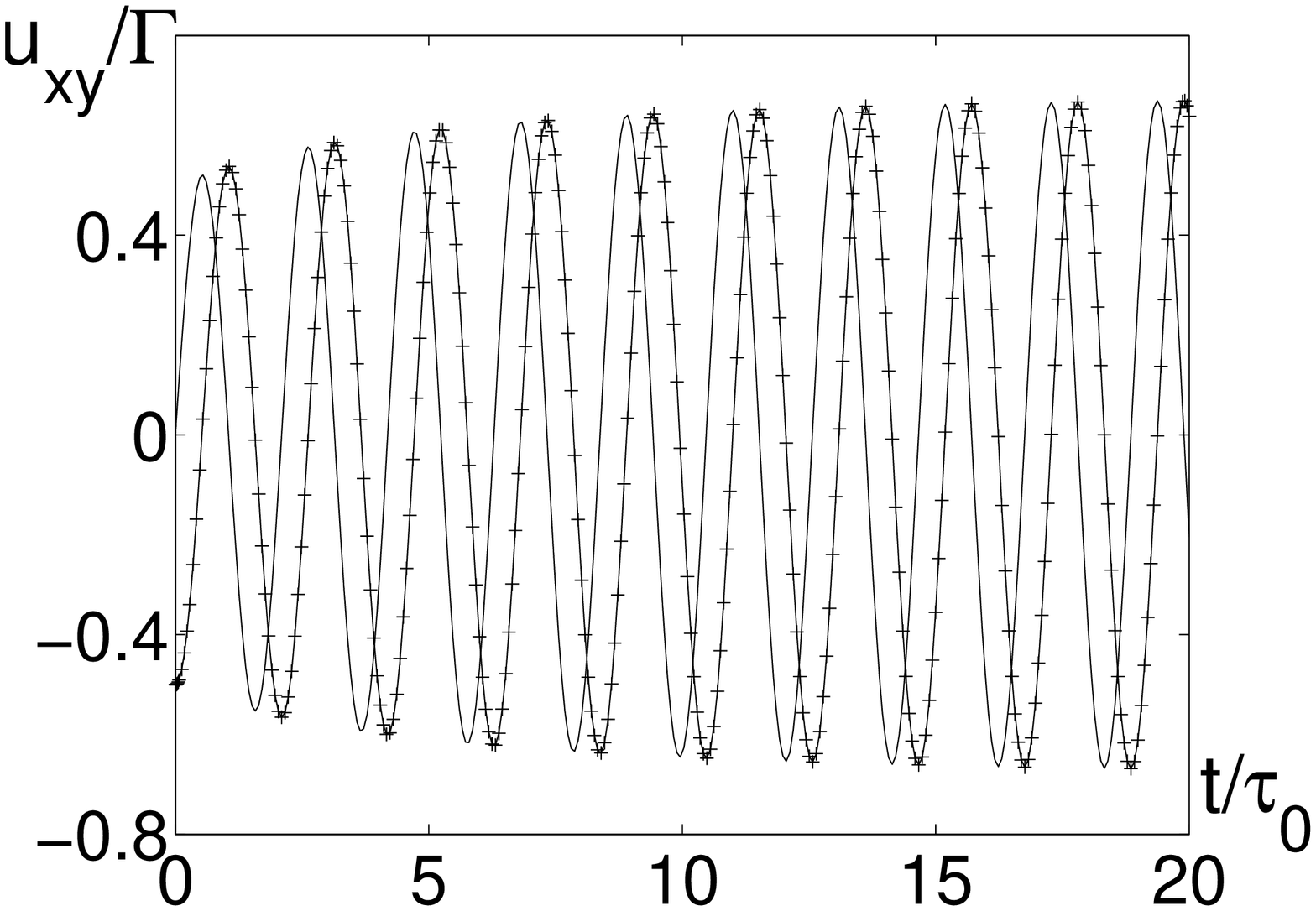}
} \caption{\scriptsize Temporal evolution of controls: the solid
line is $u_x$ and the plus-sign line is $u_y$. Here, to obtain
dimensionless quantities, $u_x$ and $u_y$ are divided by the
decoherence intensity $\Gamma$.}\label{Fig of exact decoupling for
the one-qubit systems (controls)}
\end{figure}
In the study of Lidar {\it et al} on the same problem in Ref.
\cite{Lidar}, the state variables $m_x$ and $m_y$ are exactly
decoupled from the environmental noises under the following
feedback-like control laws:
\begin{equation}\label{Control laws exactly decoupling variables from noises}
u_x=-\frac{\Gamma}{2m_z}m_y,\quad u_y=\frac{\Gamma}{2m_z}m_x,
\end{equation}
which are to be substituted into (\ref{Coherence vector equation
of the one-qubit amplitude damping decoherence model}) to get
explicit open-loop control laws. This leads to the following
equations:
\begin{eqnarray}\label{Equation of mz for exact decoupling}
\left\{%
\begin{array}{ll}
    m_x(t)=m_x^0(t)=m_{0x}\cos\omega t+m_{0y}\sin\omega t, \\
    m_y(t)=m_y^0(t)=-m_{0x}\sin\omega t+m_{0y}\cos\omega t, \\
    m_z\dot{m}_z=-\Gamma m_z^2-\Gamma m_z-\frac{\Gamma}{2}C_0^2, \\
\end{array}%
\right.
\end{eqnarray}
where $m_0=(m_{0x},m_{0y},m_{0z})^T$ and
$C_0^2=m_{0x}^2+m_{0y}^2$. Generally speaking, the last equation
of (\ref{Equation of mz for exact decoupling}) has no analytic
solutions and we can only obtain numerical solutions. With the
same parameters in the above example, we can obtain plots of
controls $u_x,\,u_y$ in Figure \ref{Fig of exact decoupling for
the one-qubit systems (controls)}.

Compare Figure \ref{Fig of the one-qubit systems (controls)} and
\ref{Fig of exact decoupling for the one-qubit systems
(controls)}, it can be shown that the control laws in Figure
\ref{Fig of exact decoupling for the one-qubit systems (controls)}
are more complex, and they approach to our laws asymptotically.
Furthermore, a divergent solution of the last equation of
(\ref{Equation of mz for exact decoupling}) may lead to divergent
control fields. With simple calculations, it can be shown that
solution of the last equation of (\ref{Equation of mz for exact
decoupling}) is convergent if and only if
\begin{eqnarray*}
m_{z_0}<\frac{-1+\sqrt{1-2C_0^2}}{2}.
\end{eqnarray*}
Figure \ref{Divergent control fields for the one-qubit systems}
shows the divergent control fields when the initial state is
chosen as $m_0=(\frac{\sqrt{2}}{2},0,\frac{\sqrt{2}}{2})^T$.
\begin{figure}[h]
\centerline{
\includegraphics[width=2.5in,height=1.5in]{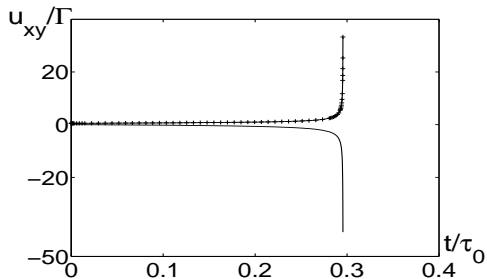}
} \caption{\scriptsize Divergent control fields for the initial
state $m_0=(\frac{\sqrt{2}}{2},0,\frac{\sqrt{2}}{2})^T$. The solid
line is $u_x$ and the plus-sign line is $u_y$. Here, to obtain
dimensionless quantities, $u_x$ and $u_y$ are divided by the
decoherence intensity $\Gamma$.}\label{Divergent control fields
for the one-qubit systems}
\end{figure}

\begin{figure}[h]
\centerline{
\includegraphics[width=2.5in,height=1.5in]{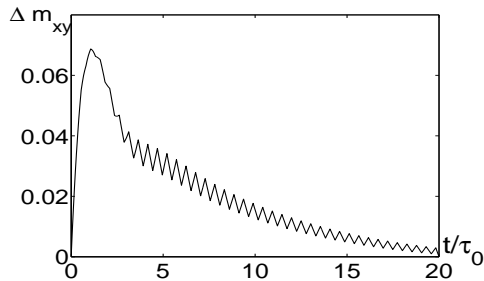}
} \caption{\scriptsize Plot of $\Delta
m_{xy}=\sqrt{(m^1_x-m^2_x)^2+(m^1_y-m^2_y)^2}$ where $m^1_x$ and
$m^1_y$ are respectively $x$, $y$ entries of the trajectory
obtained by our strategy and $m^2_x$ and $m^2_y$ are $x$, $y$
entries of the trajectory obtained by Lidar's.}\label{Fig of the
difference of m for one-qubit systems}
\end{figure}

It should be pointed out that, though the control laws obtained by
our strategy are simple and the divergence problem does not occur,
our control laws can only asymptotically, not exactly, decouple
the state variables from the environmental noises. The plot of the
difference between the trajectories obtained by our strategy and
Lidar's strategy (ideal trajectories) are shown in Figure \ref{Fig
of the difference of m for one-qubit systems}.

\subsection{Qutrit quantum systems}\label{s32}

Consider the three-level
atoms\ucite{Durt,Brukner,Grudka,Liuxiaoshu} with $\vee$
configuration as shown in Figure \ref{Fig of the three-level
atoms},
\begin{figure}[h]
\centerline{
\includegraphics[width=2.5in,height=1.6in]{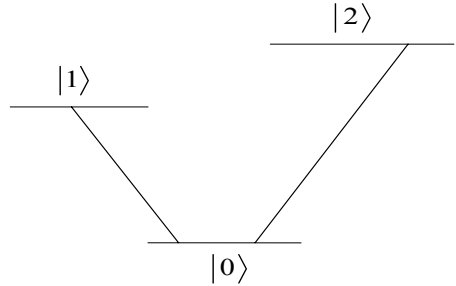}
} \caption{\scriptsize Three-level atoms with $\vee$
configuration}\label{Fig of the three-level atoms}
\end{figure}
where $|0\rangle$, $|1\rangle$ and $|2\rangle$ are the eigenstates
of the free Hamiltonian of three-level atoms with eigenvalues
$E_0<E_1<E_2$ respectively. The two excited states $|1\rangle$ and
$|2\rangle$ are coupled to the ground state $|0\rangle$.

The total Hamiltonian $H$ is expressed as $H=H_0+H_{d}$ where
$H_0$ is the free Hamiltonian and $H_{d}$ is the control
Hamiltonian. The free Hamiltonian $H_0$ can be written as:
$$H_0=E_0|0 \rangle\langle 0|+E_1|1 \rangle\langle 1 |+E_2|2 \rangle\langle 2|.$$
The control Hamiltonian
$$H_{d}=g_{10}|1\rangle\langle0|+g_{10}^*|0\rangle\langle1|+g_{20}|2\rangle\langle0|+g_{20}^*|0\rangle\langle2|$$
represents the interaction between the atoms and the driving
electromagnetic fields. The complex coefficients $g_{10},g_{20}$
can be adjusted by the amplitudes and phases of the driving
fields.

Consider the open system in which only amplitude damping
decoherence channels induced by spontaneous emissions are
introduced. The master equation model is expressed as:
\begin{equation}\label{Master equation of the three-level systems amplitude damping decoherence model}
\dot{\rho}=-i[H_0+H_d,\rho]+\Gamma_1 D[\sigma_{01}^-]\rho+\Gamma_2
D[\sigma_{02}^-]\rho,
\end{equation}
where $\sigma_{0i}^-=|0\rangle\langle i|$ is the lowering operator
from the excited state $|i\rangle$ to the ground state
$|0\rangle$. The two Lindblad terms $\Gamma_i
D[\sigma^-_{0i}]\rho$ represent the transition from $|i\rangle$ to
$|0\rangle$ caused by the spontaneous emission process.

For three-level systems, the matrix basis of the corresponding Lie
algebra $su(3)=\{-i\Omega_k\}_{k=1,\cdots,8}$ can be chosen as:
\begin{eqnarray*}
&&\Omega_1=\frac{1}{\sqrt{2}}\left(%
\begin{array}{ccc}
  0 & 0 & 0 \\
  0 & 0 & 1 \\
  0 & 1 & 0 \\
\end{array}%
\right),\,\,\Omega_2=\frac{1}{\sqrt{2}}\left(%
\begin{array}{ccc}
  0 & 0 & 0 \\
  0 & 0 & -i \\
  0 & i & 0 \\
\end{array}%
\right),\\
&&\Omega_3=\frac{1}{\sqrt{2}}\left(%
\begin{array}{ccc}
  0 & 0 & 0 \\
  0 & 1 & 0 \\
  0 & 0 & -1 \\
\end{array}%
\right),\,\,\Omega_4=\frac{1}{\sqrt{2}}\left(%
\begin{array}{ccc}
  0 & 1 & 0 \\
  1 & 0 & 0 \\
  0 & 0 & 0 \\
\end{array}%
\right),\\
&&\Omega_5=\frac{1}{\sqrt{2}}\left(%
\begin{array}{ccc}
  0 & -i & 0 \\
  i & 0 & 0 \\
  0 & 0 & 0 \\
\end{array}%
\right),\,\,\Omega_6=\frac{1}{\sqrt{2}}\left(%
\begin{array}{ccc}
  0 & 0 & 1 \\
  0 & 0 & 0 \\
  1 & 0 & 0 \\
\end{array}%
\right),\\
&&\Omega_7=\frac{1}{\sqrt{2}}\left(%
\begin{array}{ccc}
  0 & 0 & -i \\
  0 & 0 & 0 \\
  i & 0 & 0 \\
\end{array}%
\right),\,\,\Omega_8=\frac{1}{\sqrt{6}}\left(%
\begin{array}{ccc}
  -2 & 0 & 0 \\
  0 & 1 & 0 \\
  0 & 0 & 1 \\
\end{array}%
\right).
\end{eqnarray*}
It can be verified that $su(3)=p\oplus\epsilon$ is a Cartan
decomposition where
\begin{eqnarray*}
&&p=\{-i\Omega_4,\,-i\Omega_5,\,-i\Omega_6,\,-i\Omega_7\},\\
&&\epsilon=\{-i\Omega_1,\,-i\Omega_2,\,-i\Omega_3,\,-i\Omega_8\}.
\end{eqnarray*}

With the basis matrices $\{\Omega_k\}$, the free Hamiltonian can
be rewritten as:
$$H_0=\omega_3\Omega_3+\omega_8\Omega_8+\frac{E_0+E_1+E_2}{3}I,$$
where
$$\omega_3=\frac{\sqrt{2}}{2}(E_1-E_2),\,\,\omega_8=\frac{\sqrt{6}}{6}(E_1+E_2-2E_0).$$
Since the constant energy $(E_0+E_1+E_2)/3$ only contributes a
global phase to the system state, it is sufficient to consider the
following traceless free Hamiltonian
$$H_0=\omega_3\Omega_3+\omega_8\Omega_8.$$
Furthermore, let $g_{10}=\frac{1}{\sqrt{2}}(u_4+i u_5)$,
$g_{20}=\frac{1}{\sqrt{2}}(u_6+i u_7)$. The control Hamiltonian
can be written as:
$$H_d=u_4\Omega_4+u_5\Omega_5+u_6\Omega_6+u_7\Omega_7,$$
where $u_i's$ are the controls to be designed.

Now, from the coherence vector representation of $\rho$, i.e.,
\begin{equation}\label{Coherence vector representation of the density matrix of the three-level quantum systems}
\rho=\frac{1}{3}I+\sum_{i=1}^8 m_i\Omega_i,\quad
m_i=tr(\Omega_i\rho),
\end{equation} we obtain the following coherence
vector representation of the master equation (\ref{Master equation
of the three-level systems amplitude damping decoherence model}):
\begin{equation}\label{Coherence vector equation of the three-level systems amplitude damping decoherence model}
\dot{m}=\omega (\omega_3 O_3+\omega_8 O_8) m+\sum_{i=4}^7 u_i O_i
m+Dm+g,
\end{equation}
where $O_i=ad(-i\Omega_i)$. It can be verified from (\ref{Master
equation of the three-level systems amplitude damping decoherence
model}) and (\ref{Coherence vector equation of the three-level
systems amplitude damping decoherence model}) that the assumptions
(H1), (H2) and (H3) are all satisfied. According to theorem
\ref{Noise decoupling control design}, the variables
$m_i,\,i=4,5,6,7$ in the equation (\ref{Coherence vector
representation of the density matrix of the three-level quantum
systems}) can be asymptotically decoupled from the environmental
noises under the following control law:
$$u=(u_4,u_5,u_6,u_7)^T=e^{-O^{11}_0(t-t_0)}\xi,$$
where
\begin{eqnarray*}
O^{11}_0&=&\omega_3\left(%
\begin{array}{cccc}
  0 & \frac{\sqrt{2}}{2} & 0 & 0 \\
  -\frac{\sqrt{2}}{2} & 0 & 0 & 0 \\
  0 & 0 & 0 & -\frac{\sqrt{2}}{2} \\
  0 & 0 & \frac{\sqrt{2}}{2} & 0 \\
\end{array}%
\right)\\
&&+\omega_8\left(%
\begin{array}{cccc}
  0 & \frac{\sqrt{6}}{2} & 0 & 0 \\
  -\frac{\sqrt{6}}{2} & 0 & 0 & 0 \\
  0 & 0 & 0 & \frac{\sqrt{6}}{2} \\
  0 & 0 & -\frac{\sqrt{6}}{2} & 0 \\
\end{array}%
\right),
\end{eqnarray*}
and $\xi$ can be numerically solved from the equation
(\ref{Stationary algebraic equation}). Unlike one-qubit quantum
systems, the algebraic equation (\ref{Stationary algebraic
equation}) has no analytic solution in this case.

Let $\Gamma_1=\Gamma_2=\Gamma$, $t_0=0$,
$E_0=-\frac{13.6}{\tau_0}$, $E_1=--\frac{13.6}{4\tau_0}$,
$E_2=-\frac{13.6}{9\tau_0}$ and the initial state be the mixed
state:
\begin{eqnarray*}
\rho_0&=&\frac{1}{2}\left(
\frac{\sqrt{2}}{2}|0\rangle+\frac{\sqrt{2}}{2}|1\rangle
\right)\left(
\frac{\sqrt{2}}{2}\langle0|+\frac{\sqrt{2}}{2}\langle1| \right)\\
&&+\frac{1}{2}\left(
\frac{\sqrt{2}}{2}|0\rangle+\frac{\sqrt{2}}{2}|2\rangle
\right)\left(
\frac{\sqrt{2}}{2}\langle0|+\frac{\sqrt{2}}{2}\langle2| \right)\\
&=&\frac{1}{3}I+\frac{\sqrt{2}}{4}\Omega_4+\frac{\sqrt{2}}{4}\Omega_6-\frac{\sqrt{6}}{12}\Omega_8,
\end{eqnarray*}
where $\tau_0$ is a time constant which is introduced to obtain
dimensionless evolution time. For concrete systems, $\tau_0$ can
be determined by the system time scale, e.g. relaxing time of
systems. With simple calculations, it can be shown that
$u_4=-0.7063\Gamma\sin(4.4t/{\tau_0})$,
$u_5=-0.7063\Gamma\cos(4.4t/{\tau_0})$,
$u_6=-0.7063\Gamma\sin(2.5111t/{\tau_0})$,
$u_7=-0.7063\Gamma\cos(2.5111t/{\tau_0})$. Here, the amplitudes
and initial phases of the control fields are obtained by
numerically solving the equations (\ref{Stationary algebraic
equation}). Simulation results of the variables $m_i$ and controls
$u_i$, $i=4,5,6,7$ are shown in Figure \ref{Fig of the three-level
systems} and \ref{Fig of the three-level systems (controls)}.
\begin{figure}[h]
\centerline{
\includegraphics[width=1.7in,height=1.2in]{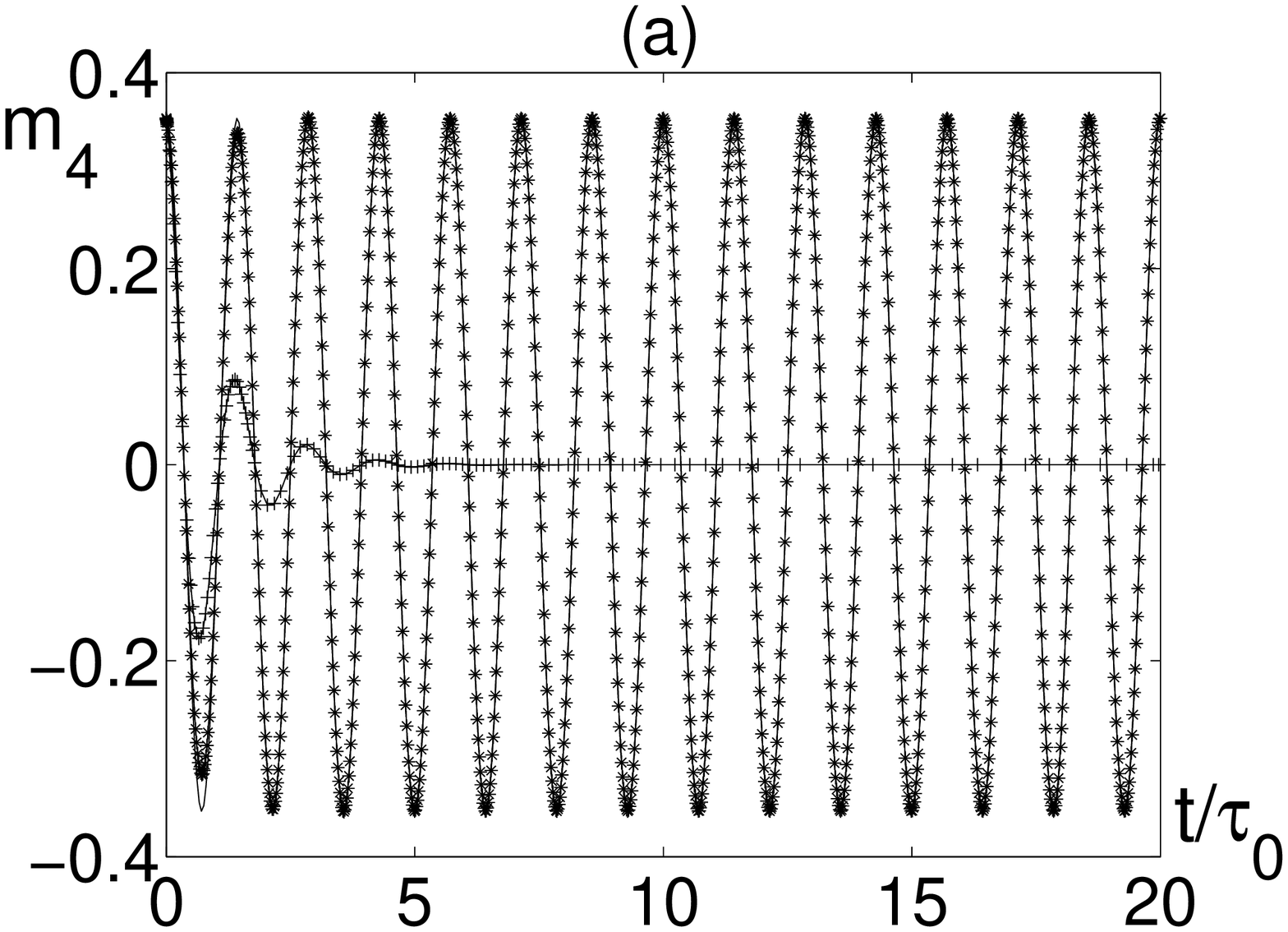}
\includegraphics[width=1.7in,height=1.2in]{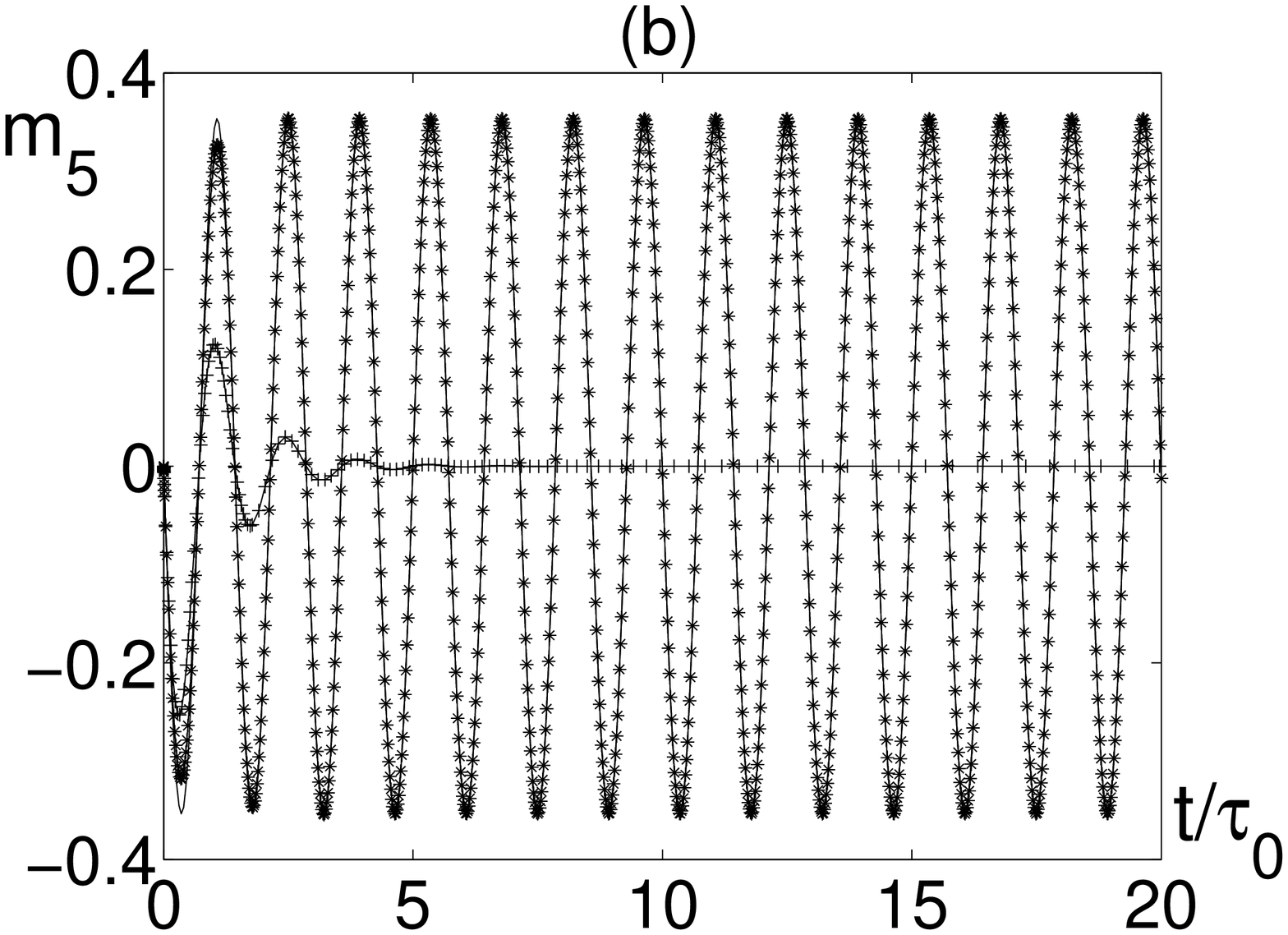}
} \centerline{
\includegraphics[width=1.7in,height=1.2in]{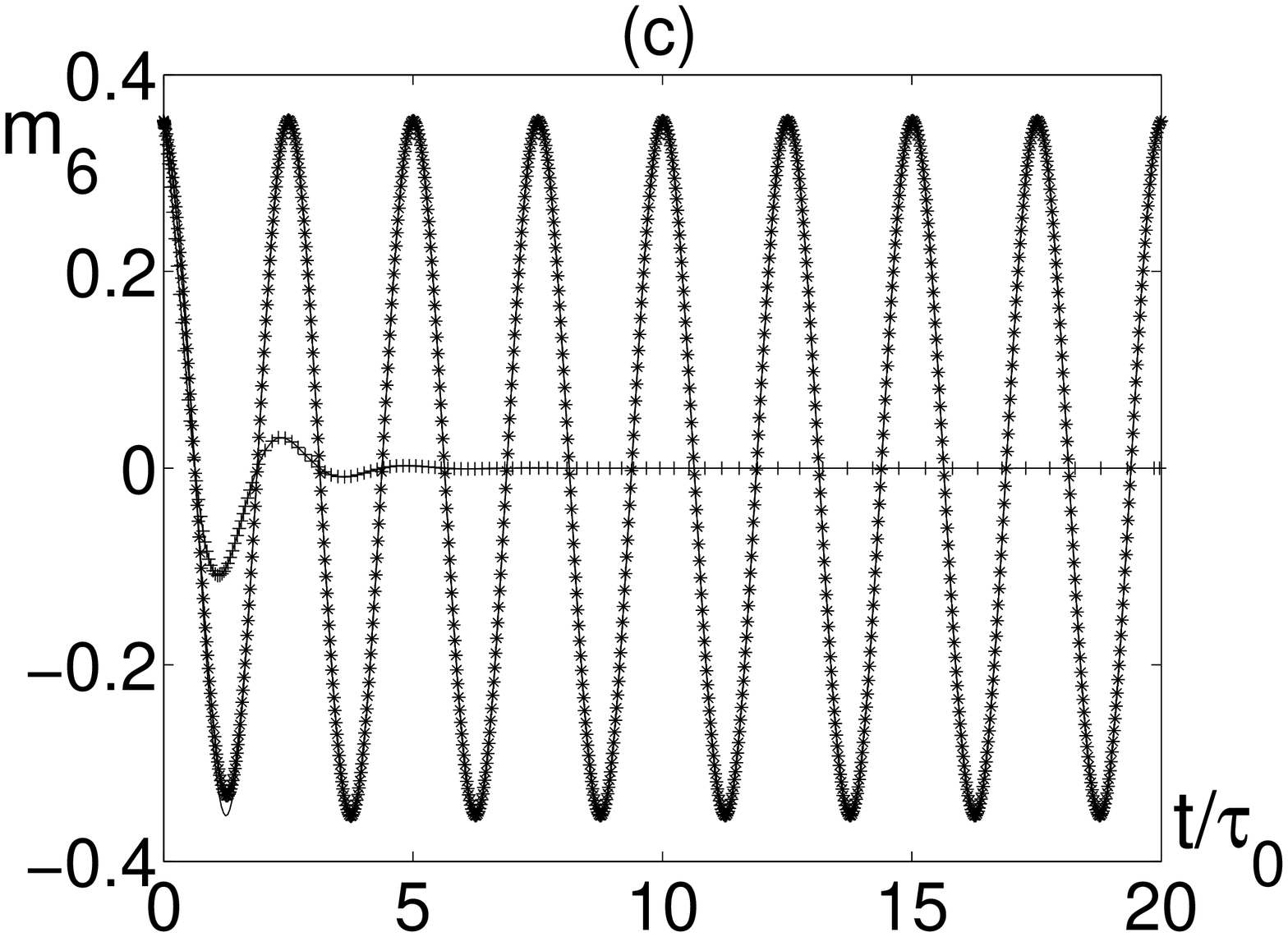}
\includegraphics[width=1.7in,height=1.2in]{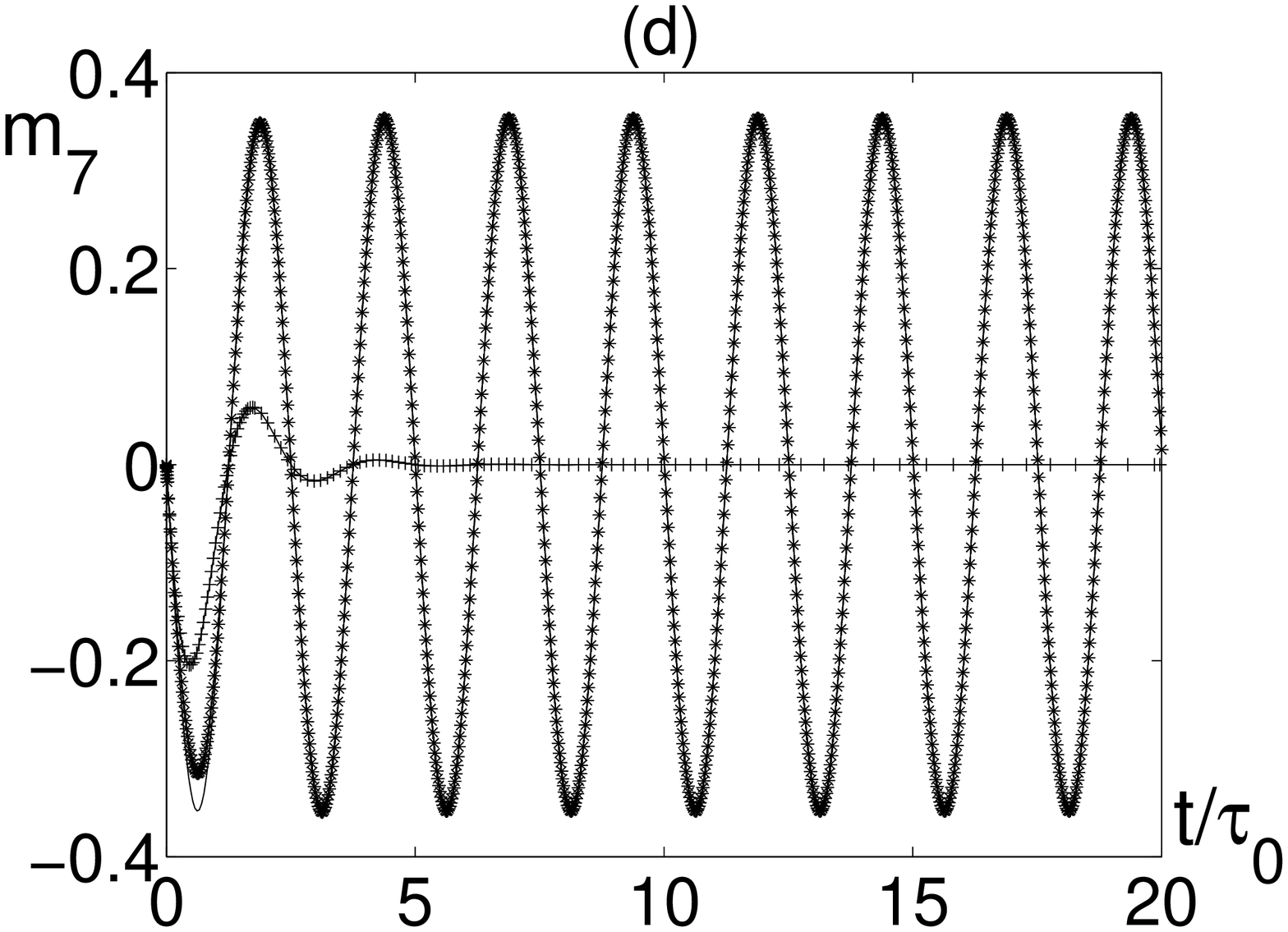}
} \caption{\scriptsize Temporal evolution of (a) $m_4$, (b) $m_5$,
(c) $m_6$ and (d) $m_7$: the asterisk line denotes the controlled
trajectory; the plus-sign line is the uncontrolled trajectory; the
solid line is the target trajectory.}\label{Fig of the three-level
systems}
\end{figure}

\begin{figure}[h]
\centerline{
\includegraphics[width=1.7in,height=1.2in]{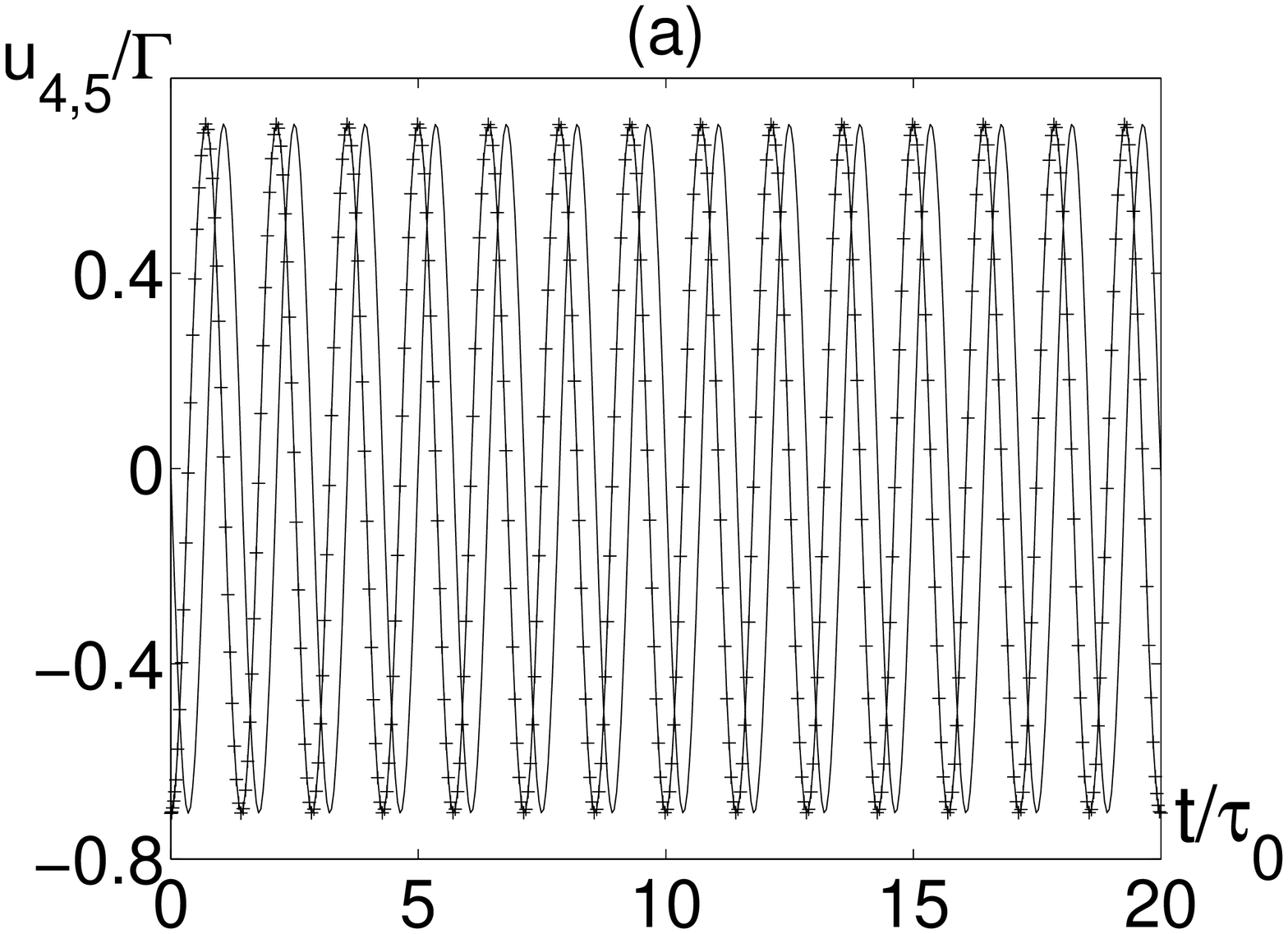}
\includegraphics[width=1.7in,height=1.2in]{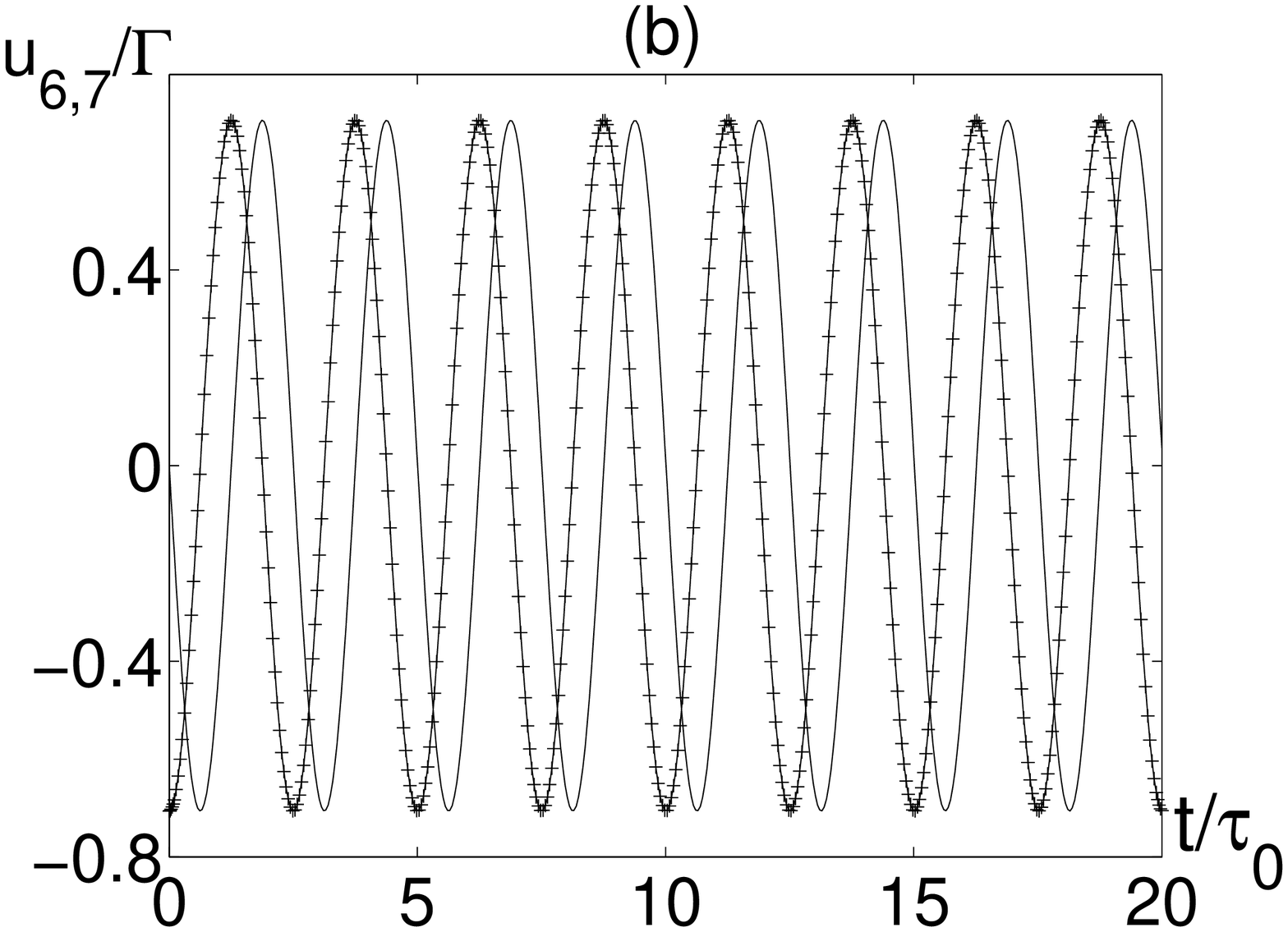}
} \caption{\scriptsize (a) Plot of $u_4$ and $u_5$ where the solid
line is $u_4$ and the plus-sign line is $u_5$; (b) Plot of $u_6$
and $u_7$ where the solid line is $u_6$ and the plus-sign line is
$u_7$. Here, to obtain dimensionless quantities, $u_i$ are divided
by the decoherence intensity $\Gamma$.}\label{Fig of the
three-level systems (controls)}
\end{figure}

Recall that the coherence $C_{01}^2=m_4^2+m_5^2$ between
$|0\rangle$ and $|1\rangle$ is determined by $m_4$ and $m_5$; the
coherence $C_{02}^2=m_6^2+m_7^2$ between $|0\rangle$ and
$|2\rangle$ is determined by $m_6$, $m_7$. The coherence in $\rho$
will vanish completely without control according to Figure
\ref{Fig of the three-level systems}. The controlled trajectory is
driven so closely to the target trajectory that they almost
coincide together, i.e., the coherence between the excited state
$|i\rangle$ and the ground state $|0\rangle$ can be asymptotically
preserved with our control strategy.

Just like one-qubit systems, if we want to exactly decouple the
corresponding variables from environmental noises, only numerical
control laws can be obtained. However, with the same parameters in
the above example, we find that the corresponding time-variant
ordinary differential equation to obtain the numerical control
laws has no solutions. In fact, in order to solve the equation, we
must calculate an inverse matrix and this matrix is singular for
the initial state in the example. It means that exactly decoupling
strategy fails for this example.

Finally, the results obtained by our control strategy can be
directly extended to the $\wedge$-type three level atoms and other
kinds of decoherence channels including phase damping decoherence
channels and depolarizing decoherence channels, as long as the
assumptions (H1), (H2) and (H3) are satisfied.

\subsection{Two-qubit quantum systems}\label{s33}

For two-qubit systems, the corresponding Lie algebra $su(4)$ has a
Cartan decomposition $su(4)=p\oplus\epsilon$, where
\begin{eqnarray*}
p&=&\{ \frac{1}{2}\sigma_i\otimes\sigma_j |i,j=x,y,z \},\\
\epsilon&=&\{ \frac{1}{2}I\otimes\sigma_j,\,
\frac{1}{2}\sigma_i\otimes I |i,j=x,y,z \}.
\end{eqnarray*}

Now the coherence vector representation of $\rho$ can be written
as:
\begin{eqnarray}\label{Two-qubit system density matrix}
\rho&=&\frac{1}{4}I\otimes I+\frac{1}{2}\sum_i
m_i^1\sigma_i\otimes I+\frac{1}{2}\sum_j m_j^2 I\otimes\sigma_j \nonumber \\
&&+\frac{1}{2}\sum_{i,j}m_{ij}^{12}\sigma_i\otimes\sigma_j,
\end{eqnarray}
where \begin{eqnarray*} &m^1_i=\frac{1}{2}tr(\sigma_i\otimes
I)\rho,\,\,m^2_j=\frac{1}{2}tr(I\otimes
\sigma_j)\rho,&\\
&m^{12}_{ij}=\frac{1}{2}tr(\sigma_i\otimes\sigma_j)\rho.&
\end{eqnarray*}
It is shown that the two-qubit variables $\{ m_{ij}^{12} \}$ can
be asymptotically decoupled from the environmental noises if the
assumptions (H1), (H2) and (H3) are all satisfied.

As an example, consider the two-qubit independent amplitude
damping decoherence model which describes two atoms that
simultaneously undergo spontaneous emissions. The control system
is described by the following master equation:
\begin{eqnarray}\label{Master equation of the two-qubit independent amplitude damping decoherence model}
\dot{\rho}&=&-i[H_0+\sum_{i,j}u_{ij}H_{ij},\rho]\nonumber\\
&&+\Gamma_1 D[\frac{1}{2}\sigma_-^1\otimes I]\rho+\Gamma_2
D[\frac{1}{2}I\otimes\sigma_-^2]\rho,
\end{eqnarray}
where $H_0=\omega_1\frac{1}{2}\sigma^1_z\otimes I+\omega_2
\frac{1}{2}I\otimes\sigma^2_z$ and
$H_{ij}=\frac{1}{2}\sigma^1_i\otimes\sigma^2_j,\,i,j=x,y,z$, are
the free and control Hamiltonians respectively;
$\sigma_-^i=\sigma_x^i-i\sigma_y^i,\,i=1,2$ is the lowering
operator of the $i^{th}$ subsystem and the positive coefficients
$\Gamma_i$ denote the corresponding decoherence rates. The two
Lindblad terms represent the amplitude damping decoherence
channels of the two subsystems.

Rewriting the master equation (\ref{Master equation of the
two-qubit independent amplitude damping decoherence model}) in the
coherence vector representation, one can verify that the system
satisfies the assumptions (H1), (H2) and (H3). Therefore, our
control strategy (\ref{Control law}) can be applied. Unlike the
one-qubit case, the quadratic algebraic equation (\ref{Stationary
algebraic equation}) does not have analytic solutions, however, we
can obtain numerical solutions of $\xi$.

Let $\omega_1=\omega_2=1/{\tau_0}$, $t_0=0$,
$\Gamma_1=\Gamma_2=\Gamma$, where $\tau_0$ is a time constant
which is introduced to obtain dimensionless evolution time, and
the initial state be the mixed state:
$$\rho_0=\frac{1}{2}\cdot\frac{1}{4}I+\frac{1}{2}|\phi_0\rangle\langle\phi_0|,$$
where $|\phi_0\rangle=\frac{1}{\sqrt{2}}(|00\rangle+|11\rangle)$
is the maximally entangled Bell state. In this case, among the
two-qubit variables $\{ m_{ij}^{12} \}$ in the equation
(\ref{Two-qubit system density matrix}), only $m^{12}_{xy}$,
$m^{12}_{yx}$, $m^{12}_{xx}$, $m^{12}_{yy}$ and $m^{12}_{zz}$ are
non-zero. The simulation results of these variables are shown in
Figure \ref{Fig of the two-qubit systems completely_controlled}.
It is shown that the uncontrolled trajectories of $m^{12}_{xy}$,
$m^{12}_{yx}$, $m^{12}_{xx}$ and $m^{12}_{yy}$ and $m^{12}_{zz}$
evolve away from anticipated values. With controls plugged in, the
controlled trajectory tracks asymptotically the target trajectory
so close that they almost coincide together.
\begin{figure}[h]
\centerline{
\includegraphics[width=1.6in,height=1.0in]{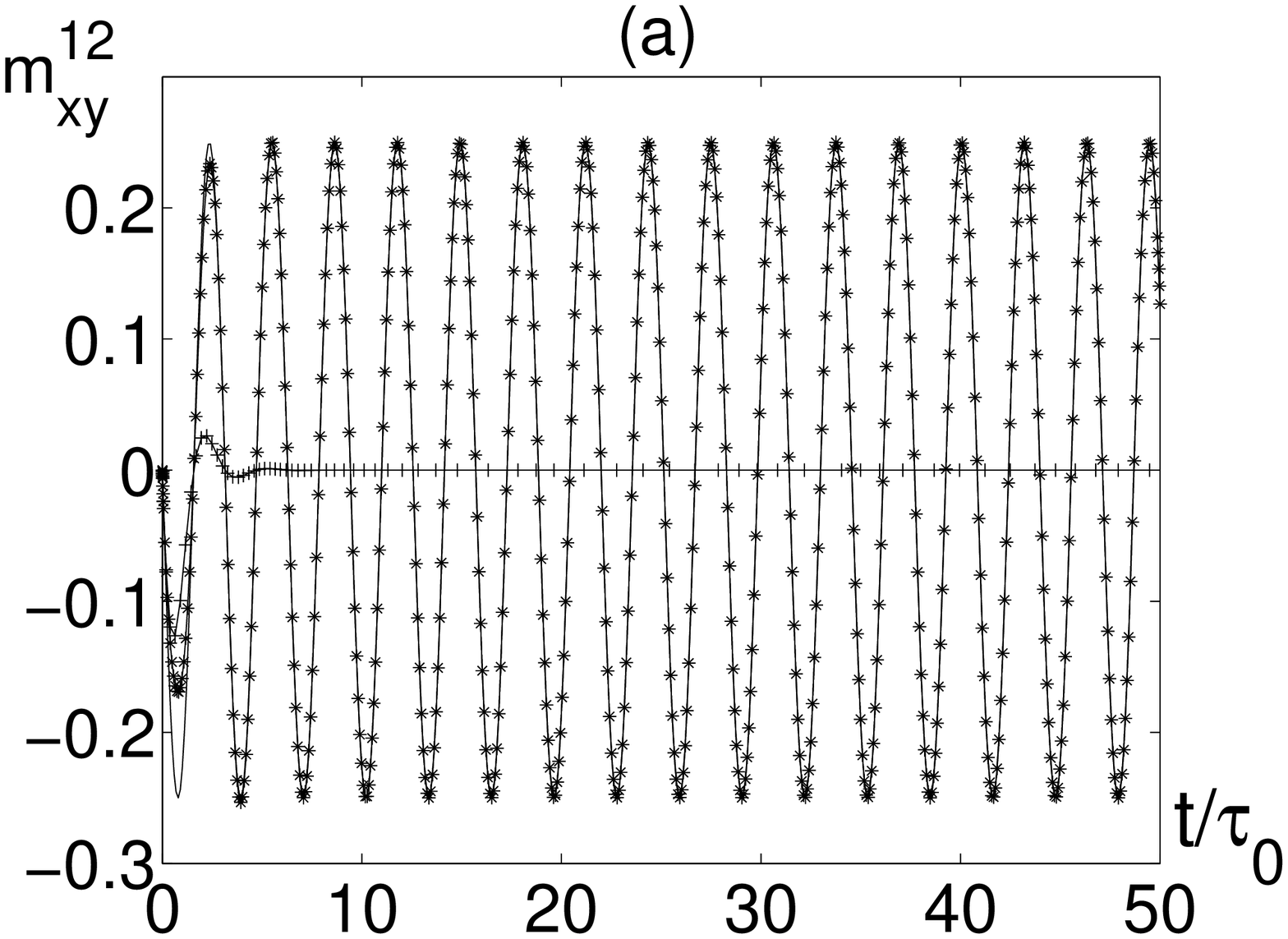}
\includegraphics[width=1.6in,height=1.0in]{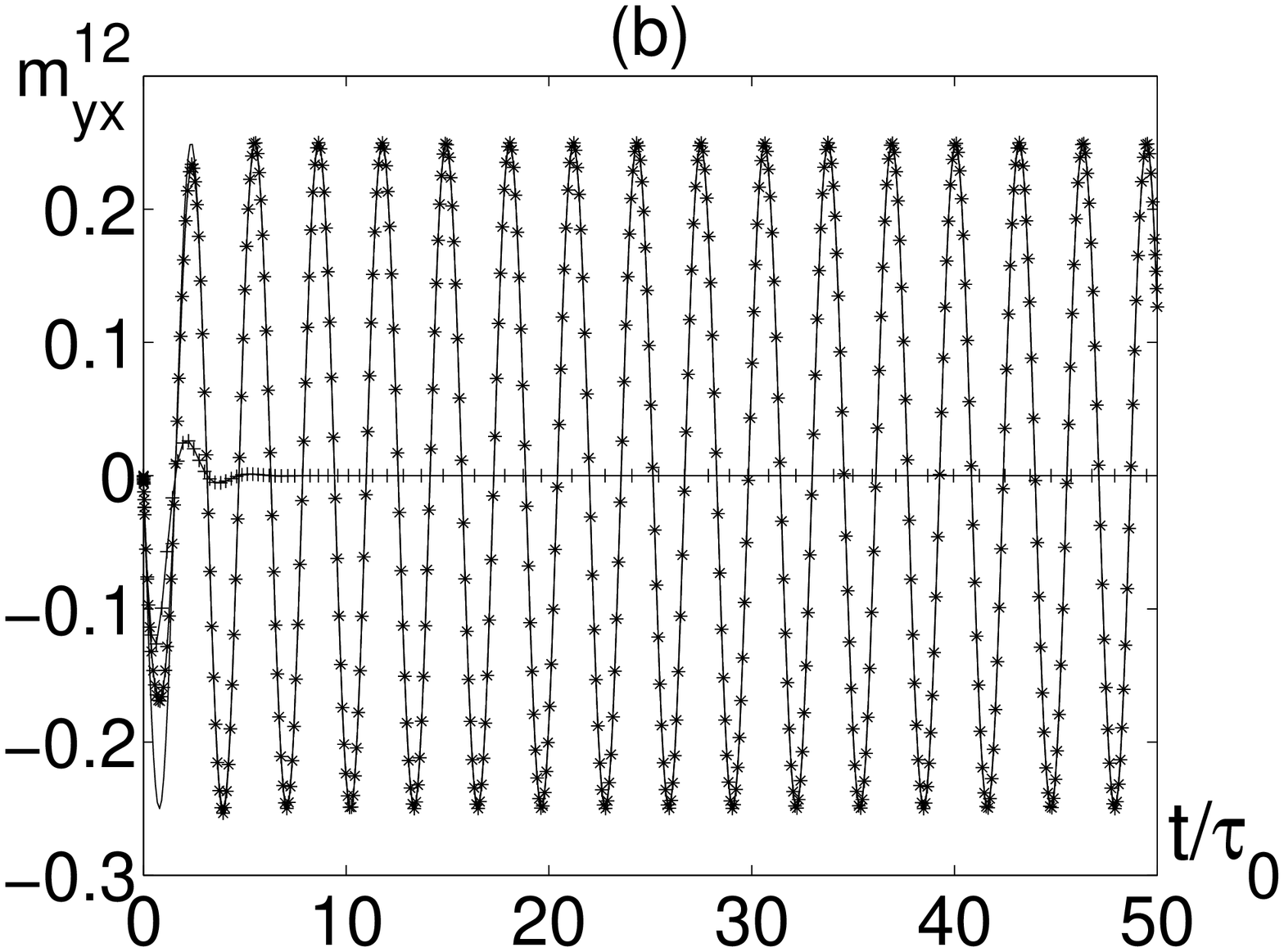}
} \centerline{
\includegraphics[width=1.6in,height=1.0in]{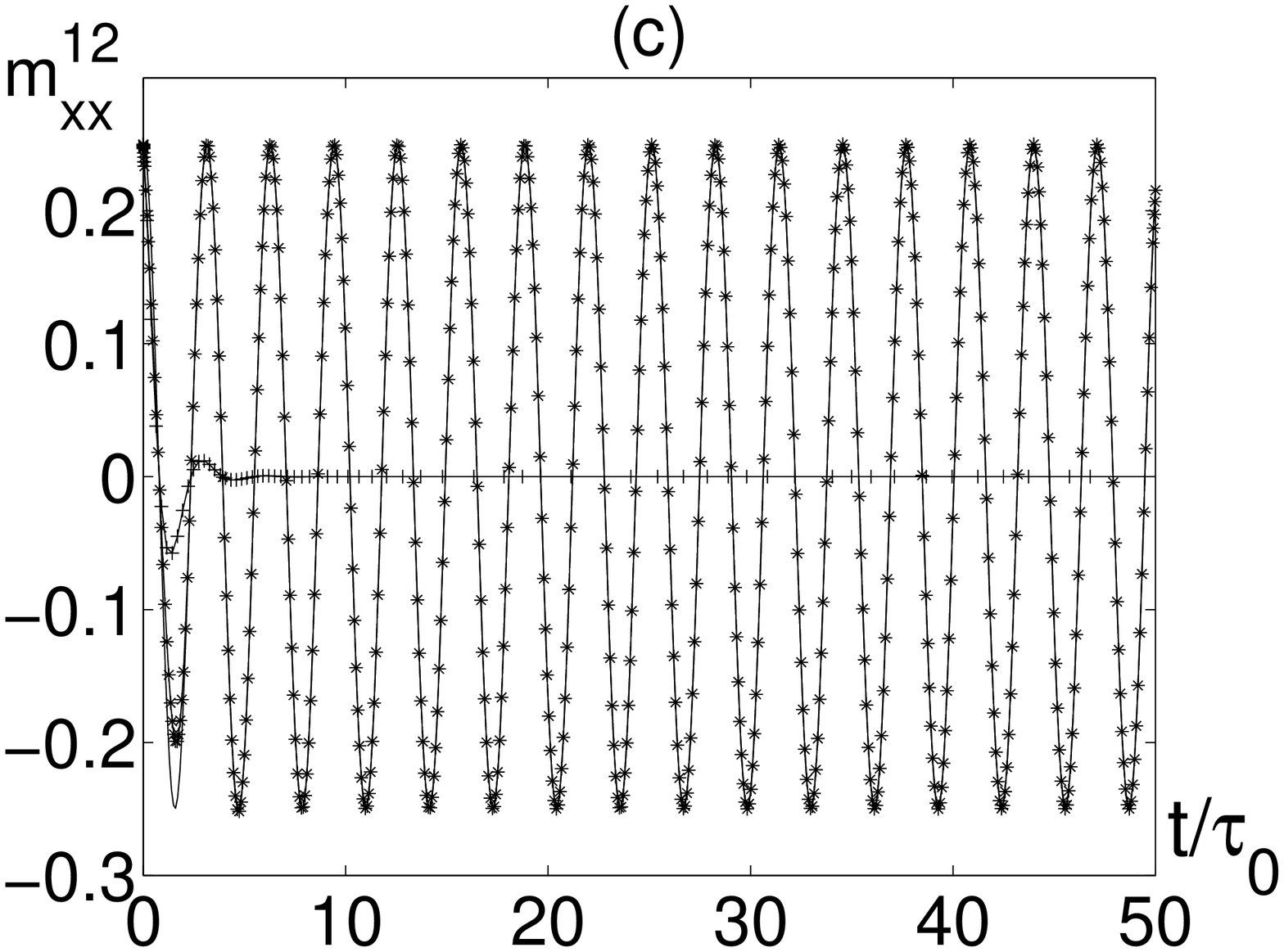}
\includegraphics[width=1.6in,height=1.0in]{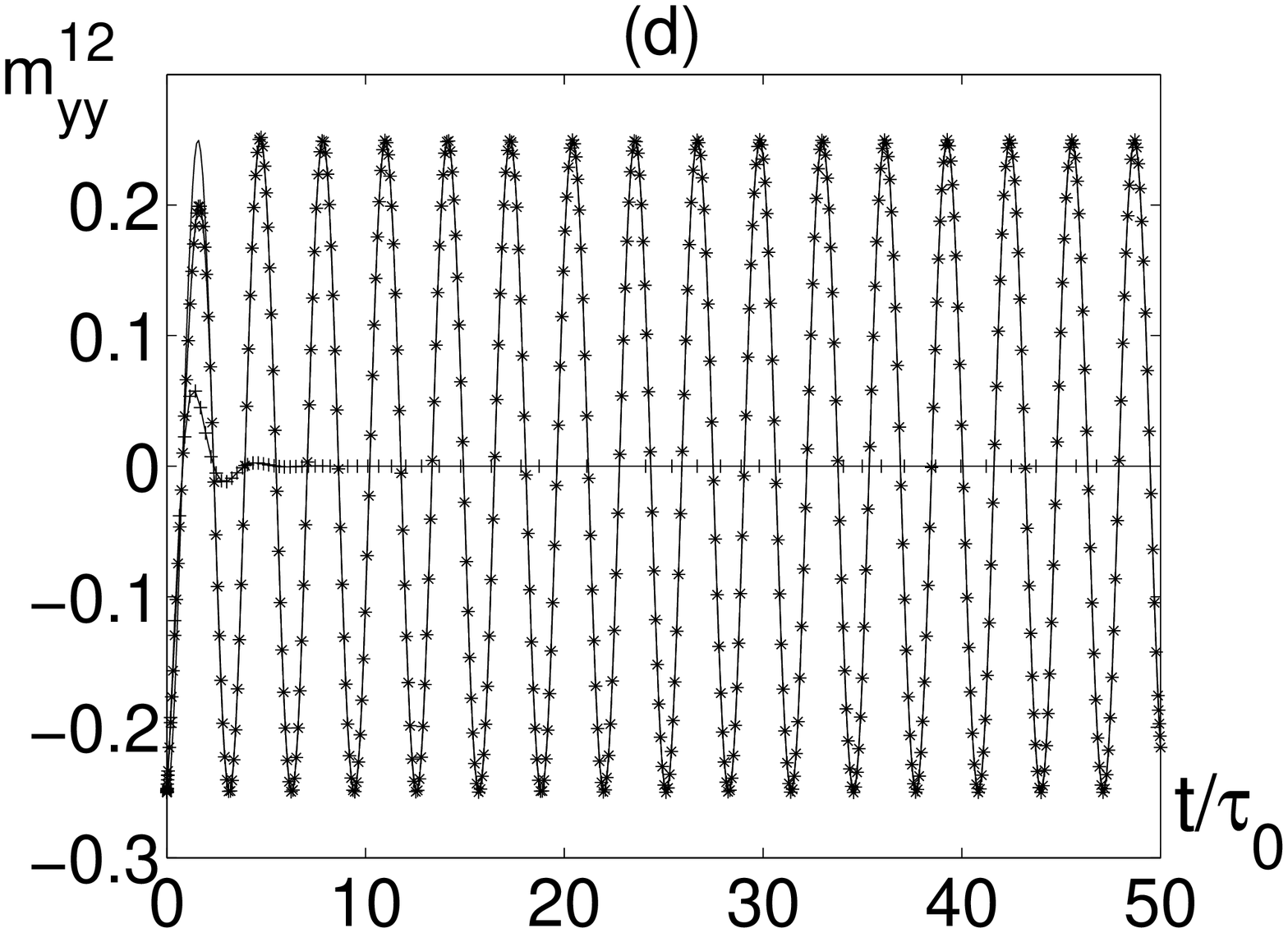}
} \centerline{
\includegraphics[width=1.6in,height=1.0in]{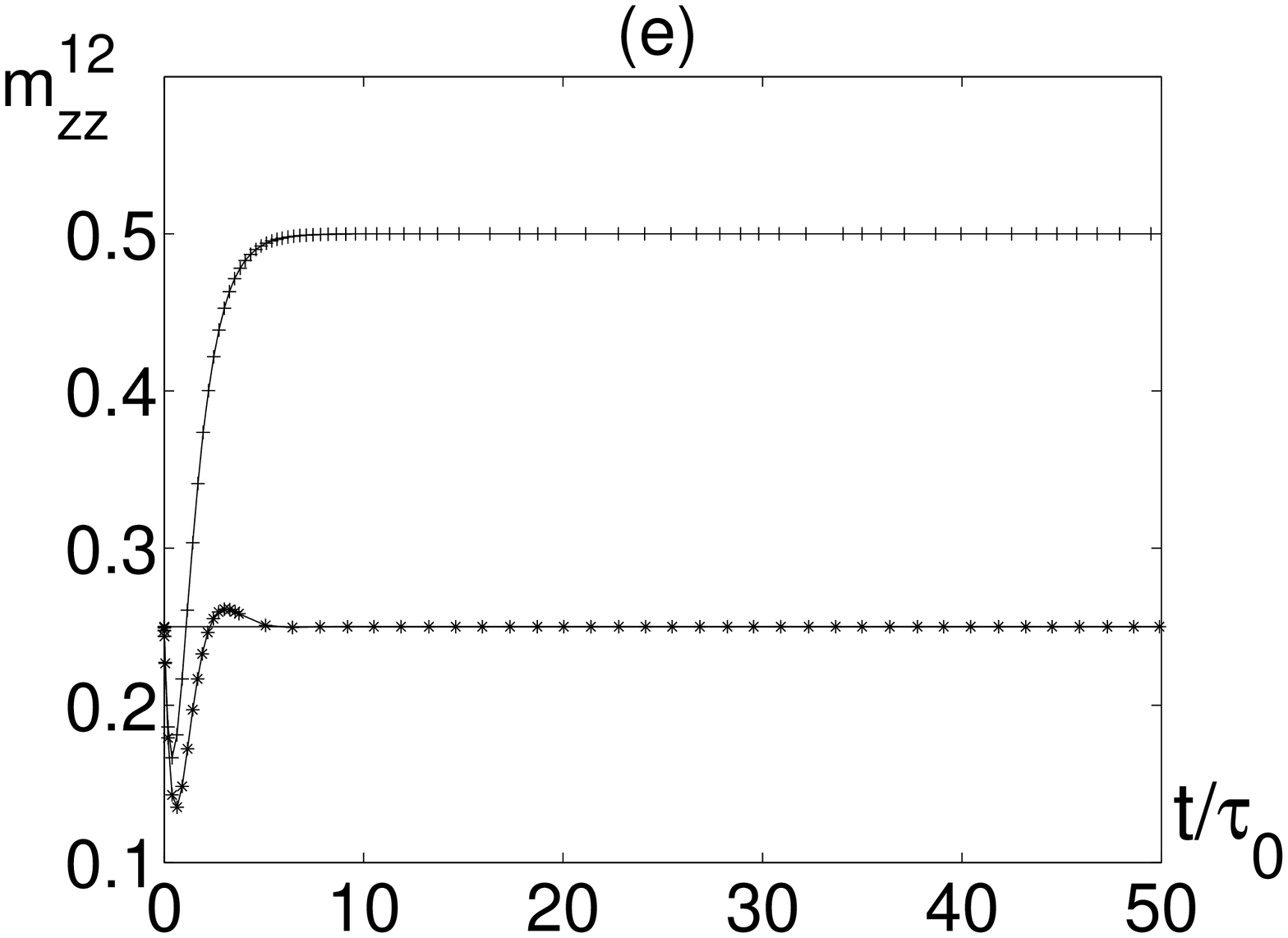}
} \caption{\scriptsize Temporal evolution of (a) $m^{12}_{xy}$,
(b) $m^{12}_{yx}$, (c) $m^{12}_{xx}$, (d) $m^{12}_{yy}$ and (e)
$m^{12}_{zz}$: the asterisk line denotes the controlled
trajectory; the plus-sign line is the uncontrolled trajectory; the
solid line is the target trajectory.}\label{Fig of the two-qubit
systems completely_controlled}
\end{figure}

Further, with simple calculations, it can be shown that only
$u_{xy}=-\frac{1}{2}\Gamma\cos(2t/{\tau_0})$,
$u_{yx}=-\frac{1}{2}\Gamma\cos(2t/{\tau_0})$,
$u_{xx}=-\frac{1}{2}\Gamma\sin(2t/{\tau_0})$,
$u_{yy}=\frac{1}{2}\Gamma\sin(2t/{\tau_0})$, and
$u_{zz}=\frac{1}{4}\Gamma$ are non-zero. Plots of the controls are
shown in Figure \ref{Fig of the two-qubit systems
completely_controlled (controls)}.
\begin{figure}[h]
\centerline{
\includegraphics[width=1.6in,height=1.0in]{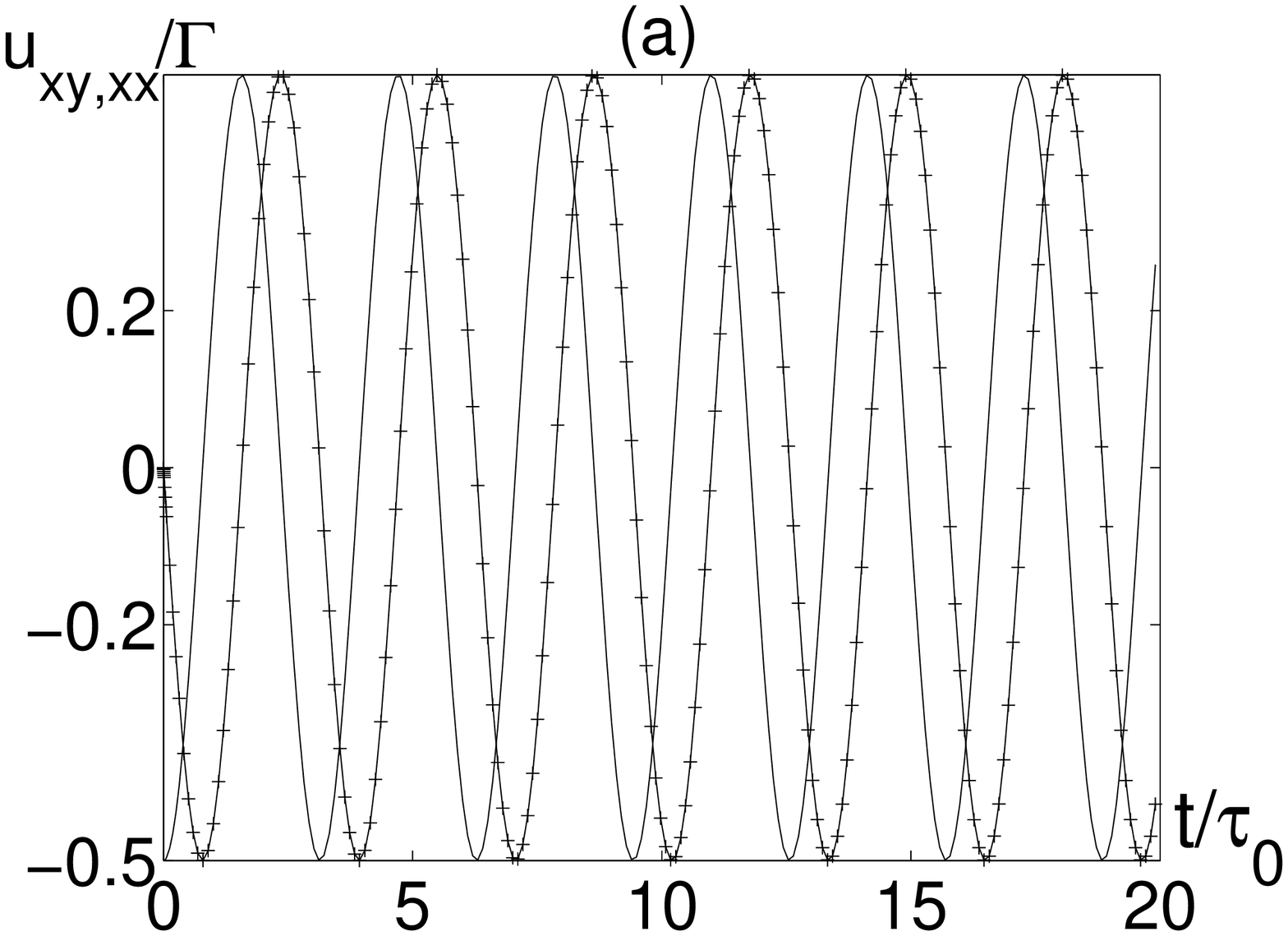}
\includegraphics[width=1.6in,height=1.0in]{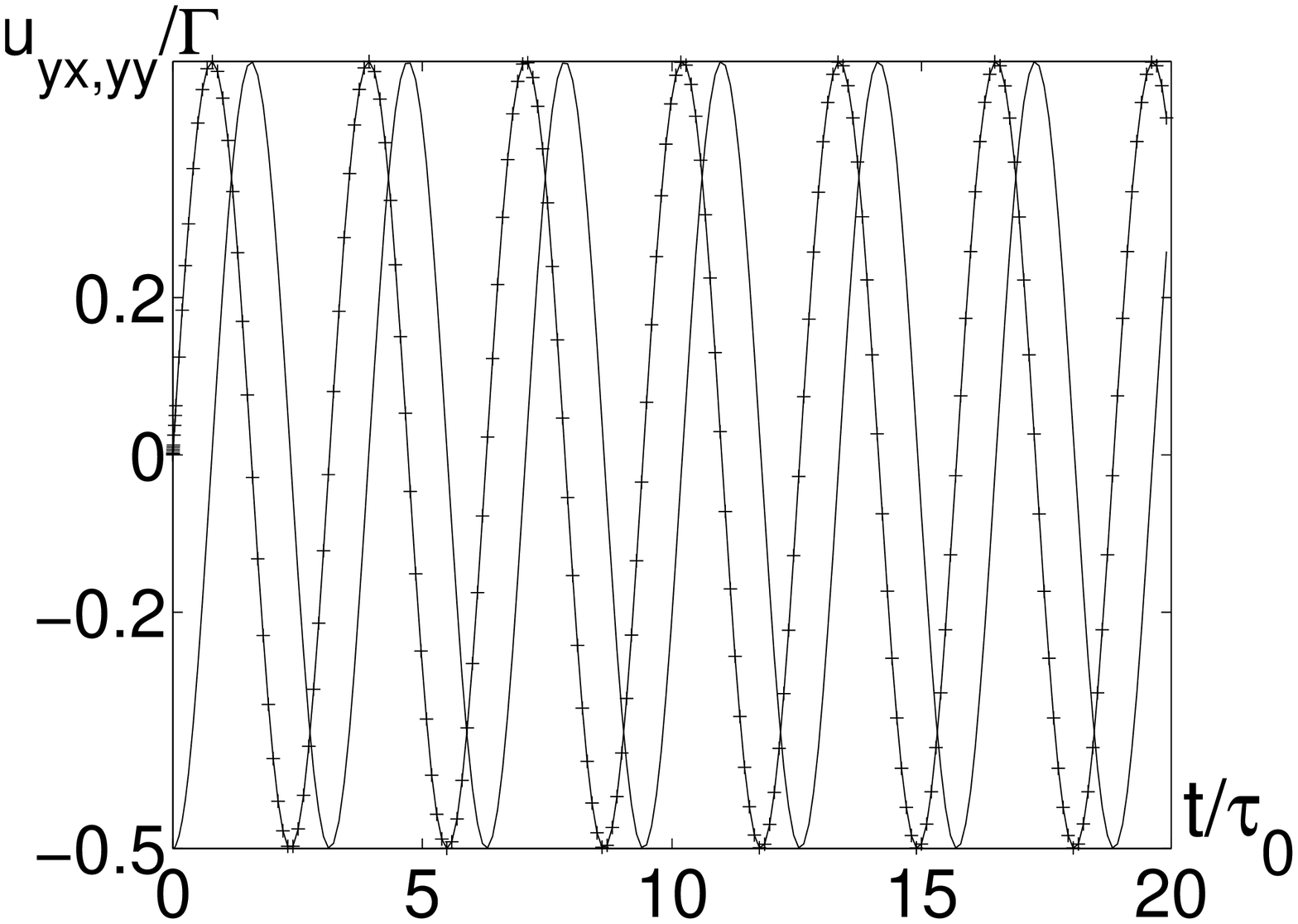}
} \caption{\scriptsize (a) Plot of $u_{xy}$ and $u_{xx}$ where the
solid line is $u_{xy}$ and the plus-sign line is $u_{xx}$; (b)
Plot of $u_{yx}$ and $u_{yy}$ where the solid line is $u_{yx}$ and
the plus-sign line is $u_{yy}$. Here, to obtain dimensionless
quantities, $u_{ij}$ are divided by the decoherence intensity
$\Gamma$.}\label{Fig of the two-qubit systems
completely_controlled (controls)}
\end{figure}

Like one-qubit and qutrit systems, if we want to exactly decouple
the variables from the environmental noises, only numerical
control laws can be obtained. With the same parameters in the
above example, we find that the corresponding time-variant
ordinary differential equation to obtain the numerical control
laws has no solutions, which means that exactly decoupling
strategy fails for this example.

It should be pointed out that the algebraic equation
(\ref{Stationary algebraic equation}) does not always have
solutions. The existence depends on the initial state. In any
case, we can always find the least-squared solution of the
algebraic equation (\ref{Stationary algebraic equation}) by
solving the following minimization problem:
$$\min_{\xi,\eta} F_1(\xi,\eta)^2+F_2(\xi,\eta)^2,$$
where $F_i(\xi,\eta),\,i=1,2$ are given in (\ref{Stationary
algebraic equation}). With such controls, we can only partially
recovers the target trajectory. See Figure \ref{Fig of the
two-qubit systems partially_controlled} for an example.

\begin{figure}[h]
\centerline{
\includegraphics[width=1.6in,height=1.0in]{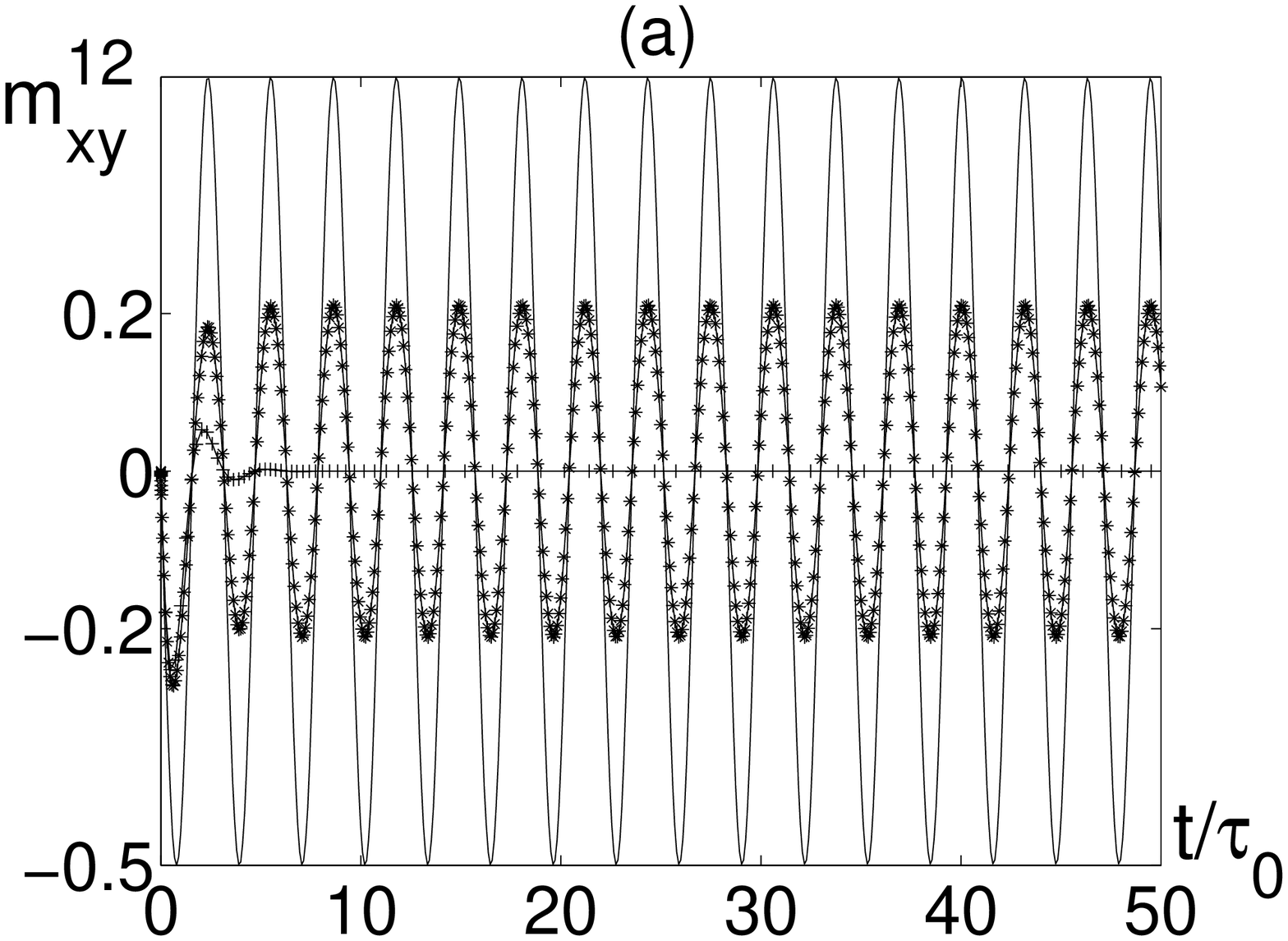}
\includegraphics[width=1.6in,height=1.0in]{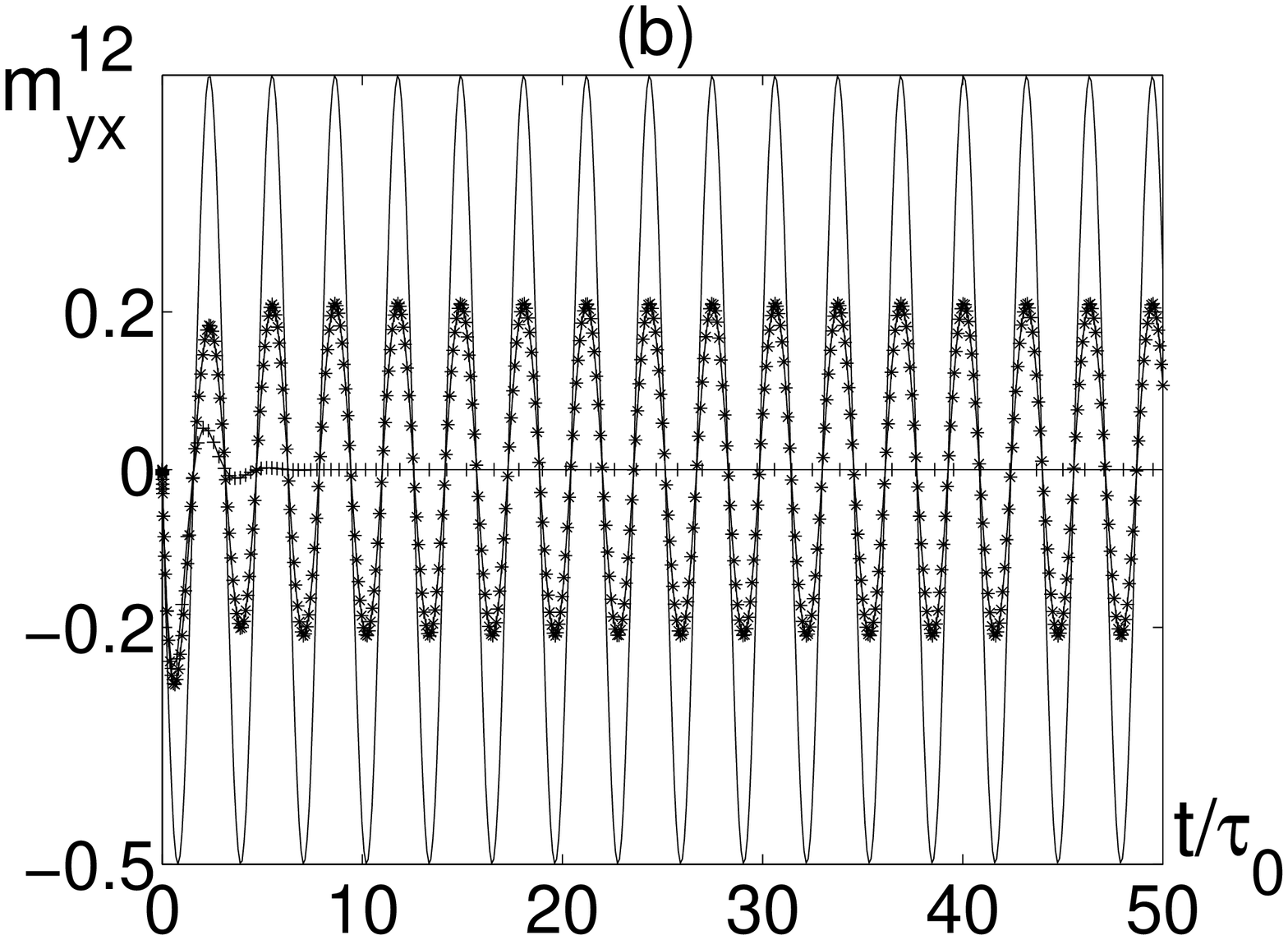}
} \centerline{
\includegraphics[width=1.6in,height=1.0in]{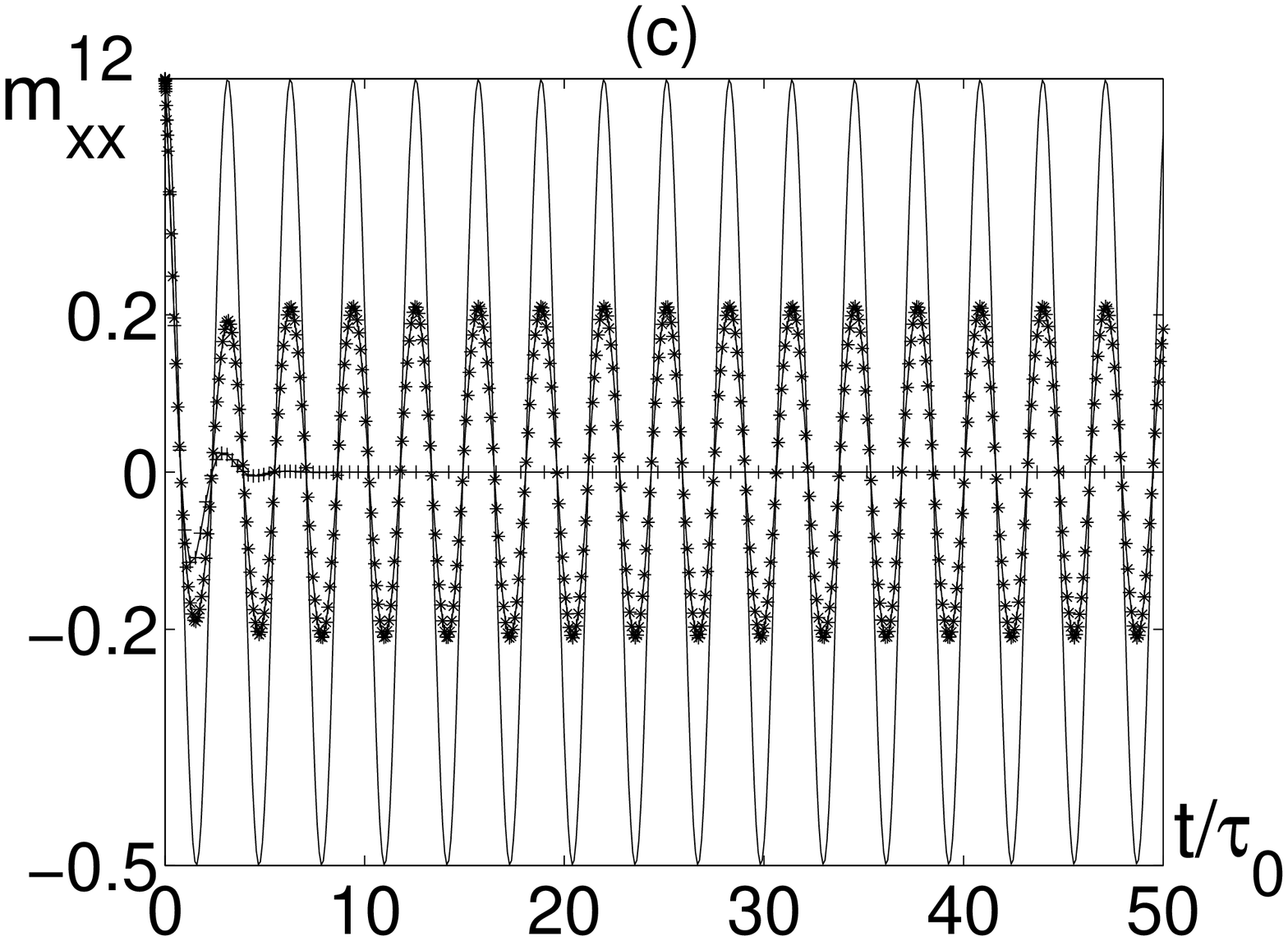}
\includegraphics[width=1.6in,height=1.0in]{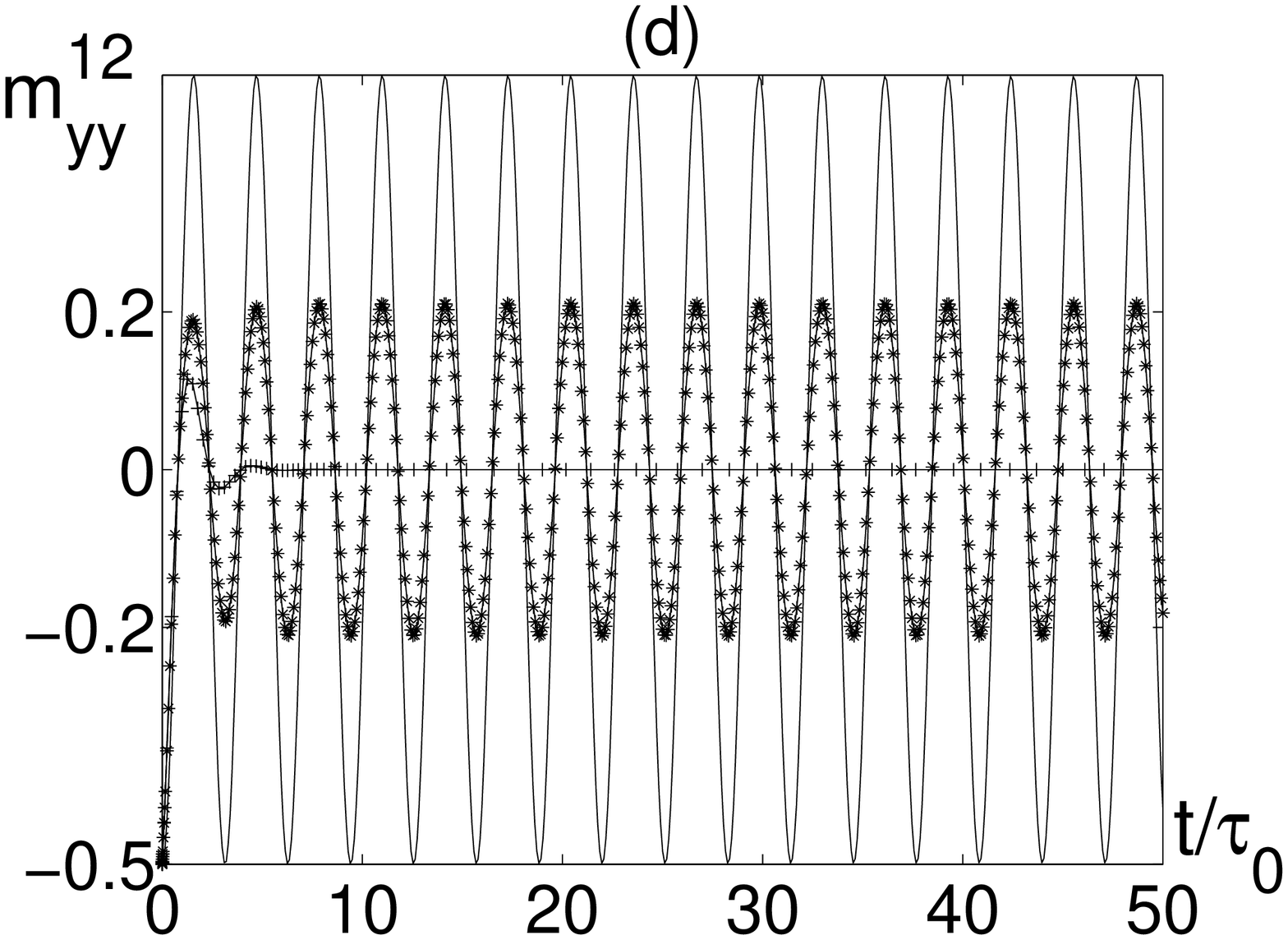}
} \centerline{
\includegraphics[width=1.6in,height=1.0in]{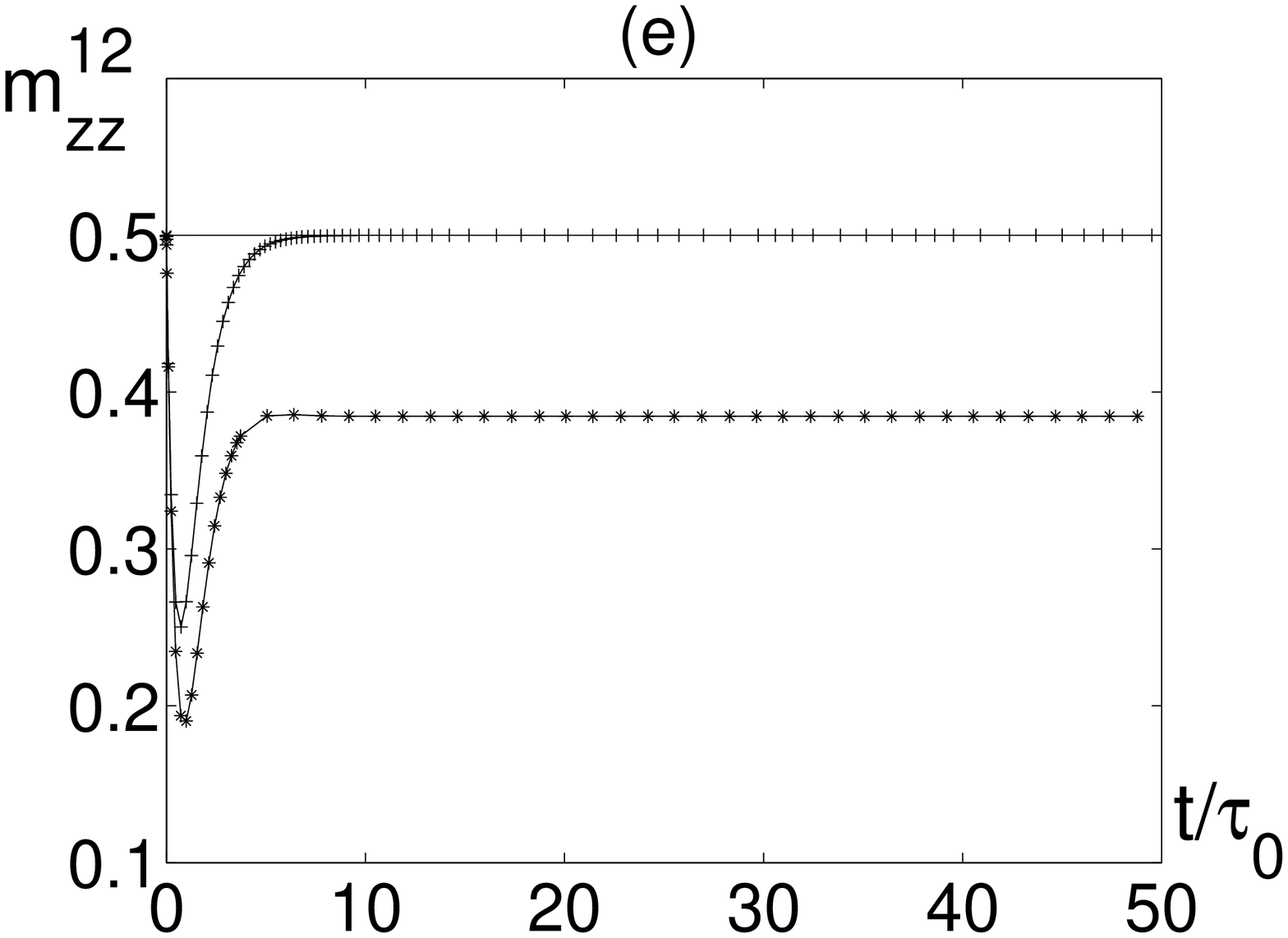}
} \caption{\scriptsize Temporal evolution of (a) $m^{12}_{xy}$,
(b) $m^{12}_{yx}$, (c) $m^{12}_{xx}$, (d) $m^{12}_{yy}$ and (e)
$m^{12}_{zz}$ with $\omega_1=\omega_2=1/{\tau_0}$,
$\Gamma_1=\Gamma_2=\Gamma$, $t_0=0$ and the initial state being
the maximal entangled Bell state
$\rho_0=|\phi_0\rangle\langle\phi_0|$, where
$|\phi_0\rangle=\frac{1}{\sqrt{2}}(|00\rangle+|11\rangle)$. The
asterisk line denotes the controlled trajectory; the plus-sign
line is the uncontrolled trajectory; the solid line is the target
trajectory.}\label{Fig of the two-qubit systems
partially_controlled}
\end{figure}

\section{Discussion}\label{s4}

In subsection \ref{s33}, it has been pointed out that the
two-qubit variables $\{ m_{ij}^{12} \}$ can be asymptotically
decoupled from the noises. In Ref. \cite{Zhangjing2}, we proposed
a multipartite mixed-state entanglement measure modified from
Jaeger's Minkowskian norm entanglement
measure\ucite{Jaeger1,Jaeger2,Jaeger3}, which, for two-qubit
states, is defined as:
$$E(\rho)=max\left\{ 2\sum_{i,j}(m_{ij}^{12})^2-\frac{1}{2},\,\,0 \right\}.$$
The entanglement measure is only related to the two-qubit
variables $\{m_{ij}^{12} \}$. For this reason, it is reasonable to
expect that the entanglement of states can be asymptotically
preserved by our control strategy.

Unfortunately, we find that, for most entangled states, the
algebraic equation (\ref{Stationary algebraic equation}) has no
solution and our control strategy can not preserve entanglement
completely under the least-squared solution. It should be further
studied to what extend our control strategy may help to preserve
entanglement.

Our asymptotical noise decoupling strategy applies control
Hamiltonians from the Cartan decomposition of the Lie algebra
$su(N)$ to decouple the systems from the environmental noises
under reasonable assumptions. Such control may not be applicable
in laboratory in present condition, especially for the two-qubit
example, but it still provides useful hints in systematic design
of decoherence control.

The construction of the Cartan decomposition is essential in our
scheme. Since the decomposition is not unique, the finding of a
"good" decomposition to achieve the expected control performance
is an interesting problem that needs further research.
\\[0.2cm]

\begin{center}
\textbf{ACKNOWLEDGMENTS}
\end{center}

This research was supported in part by the National Natural
Science Foundation of China under Grant Number 60433050, 60674039,
60635040. T.J. Tarn would also like to acknowledge partial support
from the U.S. Army Research Office under Grant W911NF-04-1-0386.
\\[0.2cm]
\appendix
\section{Proof of theorem \ref{Noise decoupling control design}}\label{Proof Of the noise decoupling control design theorem}
Before presenting the proof of the theorem, we first introduce two
lemmas:
\begin{appendixlemma}\label{Stability of the time-variant linear
systems}\ucite{Rugh} Consider the following time-variant linear
system:
\begin{equation}\label{Time-variant linear system}
\dot{x}(t)=A(t)x(t), \,\, x(t_0)=x_0.
\end{equation}
If there exist a $N\times N$ matrix $Q(t)$ and positive numbers
$\eta,\,\mu,\,\nu$ such that:
\begin{eqnarray*}
&\eta I\leq Q(t)\leq \mu I,&\\
&A^T(t)Q(t)+Q(t)A(t)+\dot{Q}(t)\leq -\nu I,&
\end{eqnarray*}
we have:
$$|x(t)|^2\leq\frac{\mu}{\eta}e^{-\frac{\nu}{\mu}(t-t_0)}|x_0|^2.$$
\end{appendixlemma}

The following lemma shows that, if the control Hamiltonians $H_i$
are chosen as $\Omega_i^p$ in (\ref{Noise decoupling decomposition
representation of the system density matrix}), the control system
(\ref{Coherent vector control system}) has a simple structure.

\begin{appendixlemma}\label{Lemma of the expression of O_0, O_i}
Let $O_0=ad(-i H_0)$, where $ad(A)$ is the adjoint representation
matrix of $A$, then

\begin{enumerate}
    \item [(1)] we have $O_0=diag(O_0^{11},O_0^{22})$, where $O_0^{11},\,O_0^{22}$ are
respectively, $m$ and $N^2-m-1$ dimensional square anti-symmetric
matrices. For $O^p_i=ad (-i\Omega_i^p)$, we have:
$$O^p_i=\left(%
\begin{array}{cc}
   & O_i^{12} \\
  -(O_i^{12})^T & \\
\end{array}%
\right).$$
    \item [(2)] we have the following equation:
    $$e^{-O_0(t-t_0)}O_i^p e^{O_0(t-t_0)}=\sum_{j=1}^m
(e^{-O^{11}_0(t-t_0)})_{ij}O_j^p,$$ where
$(e^{-O^{11}_0(t-t_0)})_{ij}$ is the $ij^{th}$ entry of the
m-dimensional matrix $e^{-O^{11}_0(t-t_0)}$.
\end{enumerate}
\end{appendixlemma}

\noindent\textbf{Proof.}

\begin{enumerate}
    \item [(1)] Corresponding to the Cartan decomposition
(\ref{Noise decoupling decomposition of su(N)}), we have the
following decomposition of the representation space
$\mathbb{R}^{N^2-1}$:
\begin{equation}\label{Decomposition of the representation space}
\mathbb{R}^{N^2-1}=\mathbb{R}^{p}\oplus\mathbb{R}^{\epsilon},
\end{equation}
where
\begin{eqnarray*}
&&\mathbb{R}^{p}=\left\{ (a_1,\cdots,a_m,0,\cdots,0)|a_i\in
\mathbb{R}
\right\},\\
&&\mathbb{R}^{\epsilon}=\left\{
(0,\cdots,0,a_{m+1},\cdots,a_{N^2-1})|a_i\in \mathbb{R} \right\}.
\end{eqnarray*}
From the assumption (H3), $O_0$ can be written as:
\begin{equation}\label{Expression of O_0}
O_0=\sum_{l=m+1}^{N^2-1}\omega_l ad(-i \Omega_l^{\epsilon}).
\end{equation}
According to the Cartan decomposition (\ref{Noise decoupling
decomposition of su(N)}), the special structure of $O_0$ and
$O^p_i$ can be easily verified from the fact that $ad(p)$ maps
$\mathbb{R}^{p}$ into $\mathbb{R}^{\epsilon}$,
$\mathbb{R}^{\epsilon}$ into $\mathbb{R}^{p}$, and $ad(\epsilon)$
maps $\mathbb{R}^{\epsilon}$ into $\mathbb{R}^{\epsilon}$,
$\mathbb{R}^{p}$ into $\mathbb{R}^{p}$.
  \item [(2)] From (\ref{Expression of O_0}) and
the equality $ad([A,B])=[ad(A),ad(B)]$, it can be shown that:
\begin{equation}\label{Commutation of O}
[O_0,O^p_i]=\sum_{j=1}^m (O^{11}_0)_{ij}O^p_j.
\end{equation}
In fact, it can be deduced that:
\begin{eqnarray*}
&&[O_0,O_i^p]=\sum_{l=m+1}^{N^2-1}\omega_l
[ad(-i\Omega_l^{\epsilon}),ad(-i\Omega_i^p)]\\
&=&\sum_{l=m+1}^{N^2-1}\omega_l
ad([-i\Omega_l^{\epsilon},-i\Omega_i^p])=\sum_{l=m+1}^{N^2-1}\sum_{j=1}^m\omega_l
c_{lij}ad(-i\Omega_j^p)\\
&=&\sum_{j=1}^m\left(\sum_{l=m+1}^{N^2-1}\omega_l
c_{lij}\right)O_j^p=\sum_{j=1}^m (O^{11}_0)_{ij}O_j^p,
\end{eqnarray*}
where $\{c_{ijk}\}$ is the structure coefficients of the Lie
algebra $su(N)$. From (\ref{Commutation of O}), it can be easily
verified by induction that
$$[(-O_0)^{(k)},O^p_i]=\sum_{j=1}^m ((-O^{11}_0)^k)_{ij} O^p_j,$$
where $[A^{(i)},B]=[A,[A^{(i-1)},B]]$ and $[A^{(0)},B]=B$. Now,
from the equality
$$e^A B
e^{-A}=\sum_{i=0}^{\infty}\frac{1}{i!}[A^{(i)},B],$$ we have:
\begin{eqnarray*}
&& e^{-O_0(t-t_0)}O^p_i
e^{O_0(t-t_0)}\\
&=&\sum_{k=0}^{\infty}\frac{1}{k!}[(-O_0(t-t_0))^{(k)},O^p_i]\\
&=&\sum_{k=0}^{\infty}\sum_{j=1}^m\frac{1}{k!}((-O^{11}_0(t-t_0))^k)_{ij}
O^p_j\\
&=&\sum_{j=1}^m(e^{-O^{11}_0(t-t_0)})_{ij} O^p_j.
\end{eqnarray*}
$\qquad\qquad\qquad\qquad\qquad\qquad\qquad\qquad\qquad\qquad
\blacksquare$
\end{enumerate}

\noindent\textbf{Proof Of Theorem \ref{Noise decoupling control
design}.}

Substitute the control laws (\ref{Control law}) into the equation:
\begin{equation}\label{Coherent vector control system without the initial value}
\dot{m}(t)=O_0 m(t)+\sum\limits_{i=1}^{m} u_i O^p_i m(t)+Dm(t)+g.
\end{equation}
Let $m^1(t),\,m^2(t)$ are two solutions of the equation
(\ref{Coherent vector control system without the initial value})
with the initial values to be $m_0^1,\,m_0^2$ respectively, then
$m^1(t)-m^2(t)$ satisfies the following time-variant linear
equation:
\begin{equation}\label{Special time-variant linear system}
\dot{x}(t)=(D+O_0+\sum\limits_{i=1}^{m} u_i O^p_i)x(t)
\end{equation}
with the initial value $x(t_0)=m_0^1-m_0^2$.

Let $Q=I,\,\eta=\mu=1,\,\nu=2d_{min}$, where $-d_{min}<0$ (from
the assumption (H2)) is the maximal eigenvalue of $D$. From lemma
\ref{Stability of the time-variant linear systems}, it can be
verified that:
$$|m^1(t)-m^2(t)|\leq e^{-2d_{min}(t-t_0)}|m_0^1-m_0^2|,$$
which means $m^1(t)-m^2(t)\rightarrow 0$ when $t$ tends to
infinity.

From the above analysis, to show that the solution of equation
(\ref{Coherent vector control system}) tends to (\ref{Stationary
solution}), it is sufficient to prove that (\ref{Stationary
solution}) satisfies the equation (\ref{Coherent vector control
system without the initial value}), which is equivalent to
$$\sum_{i=1}^m u_i O^p_i e^{O_0(t-t_0)}m_0^{\infty}+De^{O_0(t-t_0)}m_0^{\infty}+g=0,$$
where $m_0^{\infty}=(m_0^1,\eta)^T$. From the complete decoherence
condition (H1), we have $[e^{O_0(t-t_0)},D]=0$ and
$e^{O_0(t-t_0)}g=g$, which results in
\begin{equation}\label{Temp equation}
\sum_{i=1}^m u_i O_i^p e^{O_0(t-t_0)}m_0^{\infty}+e^{O_0(t-t_0)}(D
m_0^{\infty}+g)=0.
\end{equation}
Substituting (\ref{Control law}) into the first term in (\ref{Temp
equation}) and from the lemma \ref{Lemma of the expression of O_0,
O_i}, we have:
\begin{eqnarray*}
&&(O^p_1 m^{\infty}(t),\cdots,O^p_m m^{\infty}(t))\left(
\begin{array}{c}
u_1 \\ \vdots \\ u_m \\
\end{array} \right)
\end{eqnarray*}
\begin{eqnarray*}
&=& e^{O_0(t-t_0)}\left(e^{-O_0(t-t_0)}O^p_1
e^{O_0(t-t_0)}m^{\infty}_0,\cdots,\right. \\
&&\left. e^{-O_0(t-t_0)}O^p_m e^{O_0(t-t_0)}m^{\infty}_0
\right)e^{-O^{11}_0(t-t_0)}\xi\\
&=&
e^{O_0(t-t_0)}\left(\sum_{k=1}^m(e^{-O^{11}_0(t-t_0)})_{1,k}O^p_k
m^{\infty}_0,\cdots,\right. \\
&&\left. \sum_{k=1}^m(e^{-O^{11}_0(t-t_0)})_{m,k}O^p_k m^{\infty}_0 \right)e^{-O^{11}_0(t-t_0)}\xi\\
&=& e^{O_0(t-t_0)}(O^p_1 m^{\infty}_0,\cdots,O^p_m
m^{\infty}_0)(e^{-O^{11}_0(t-t_0)})^T e^{-O^{11}_0(t-t_0)}\xi
\\
&=& e^{O_0(t-t_0)}(O^p_1 m^{\infty}_0,\cdots,O^p_m
m^{\infty}_0)e^{O^{11}_0(t-t_0)} e^{-O^{11}_0(t-t_0)}\xi
\\
&=&e^{O_0(t-t_0)}\sum_{i=1}^m \xi_i O^p_i m^{\infty}_0.
\end{eqnarray*}
Therefore, the equation (\ref{Temp equation}) is reduced to
$$e^{O_0(t-t_0)}\left( \sum_{i=1}^m \xi_i O^p_i m^{\infty}_0+D m^{\infty}_0+g \right)=0,$$
which is equivalent to (\ref{Stationary algebraic equation}).
$\qquad\qquad\qquad\qquad\qquad \blacksquare$
\\[0.2cm]

\end{document}